\documentclass[final,authoryear,5p,times,twocolumn]{elsarticle}

\usepackage{graphicx}
\usepackage{amssymb}
\usepackage{amsmath}

\usepackage{enumerate}
\usepackage{color}

\journal{Journal of Theoretical Biology}

\DeclareMathOperator{\E}{E}
\DeclareMathOperator{\Var}{Var}

\begin{document}

\begin{frontmatter}

\bibliographystyle{elsarticle-harv}

\title{On the coexistence of cooperators, defectors and conditional
cooperators in the multiplayer iterated Prisoner's Dilemma}

\author[1]{Jelena Gruji\'c}
\ead{jgrujic@math.uc3m.es}
\author[1]{Jos\'e A.\ Cuesta\corref{c1}}
\ead{cuesta@math.uc3m.es}
\ead[url]{http://gisc.uc3m.es/~cuesta}
\author[1,2]{Angel S\'anchez}
\ead{anxo@math.uc3m.es}
\ead[url]{http://allariz.uc3m.es/~anxosanchez/}

\address[1]{Grupo Interdisciplinar de Sistemas Complejos (GISC),
Departamento de Matem\'aticas, Universidad Carlos III de Madrid,
Avenida de la Universidad 30,\\ 28911 Legan\'es, Madrid, Spain}

\address[2]{Instituto de Biocomputaci\'on y F\'\i sica de Sistemas Complejos (BIFI),
Universidad de Zaragoza, Campus R\'\i o Ebro, 50018 Zaragoza, Spain}

\cortext[c1]{Corresponding author}




\begin{abstract}
Recent experimental evidence [Gruji\'c {\em et al.}, PLoS ONE {\bf 5},
e13749 (2010)] on the spatial Prisoner's Dilemma suggests that players
choosing to cooperate or not on the basis of their previous action and
the actions of their neighbors coexist with steady defectors and cooperators.
We here study the coexistence of these three strategies in the multiplayer
iterated Prisoner's Dilemma by means of the replicator dynamics. 
We consider groups with $n=2, 3, 4$ and $5$ players and compute the payoffs
to every type of player as the limit of a Markov chain where the transition
probabilities between actions are found from the corresponding strategies.
We show that for group sizes up to $n=4$ there exists an interior point
in which the three strategies coexist, the corresponding basin of attraction
decreasing with increasing number of players, whereas we have not been able to locate
such a point for $n=5$. We analytically show that in the limit $n\to\infty$ no interior 
points can arise. We conclude by discussing the implications of this theoretical 
approach on the behavior observed in experiments. \end{abstract}


\end{frontmatter}

\section{Introduction}

In the past few years, different mechanisms have been proposed to explain the origin and stability of cooperation \citep{nowak:2006b}.
One of these mechanisms involves assortment of cooperators
\citep{fletcher:2009}, in particular through the existence
of a spatial or social structure dictating who interacts with whom (cf.\ network 
reciprocity in \cite{nowak:2006b}). Cooperators might then interact mainly with
each other and keep the benefits of cooperation to the extent that they perform 
better than defectors or free riders in peripheral positions. This idea stems
from the work by \cite{nowak:1992}, who carried out a simulation of the iterated
Prisoner's Dilemma (PD) \citep{rapoport:1966,axelrod:1981} on a lattice in which
every individual interacted with her eight nearest neighbors. Their finding of
sizable proportions of cooperative actions even when the temptation to defect
was quite large stimulated a large amount of work on evolutionary game theory
on graphs (for reviews see, e.g., \cite{szabo:2007} and \cite{roca:2009a}).
Unfortunately, in spite of the large body of theoretical work devoted to this
issue, it has not been possible to 
reach a general conclusion about how the existence of structure on a population
could promote cooperation: indeed, it was shown that the emergence and 
survival of cooperative behaviors depended so crucially on the details of the
models that their applicability to real life situations was dubious, at best. 

In view of this situation, in the last few years a number of groups have carried
out experiments to probe the relationship between population structure and 
cooperation with real human subjects
\citep{kirchkamp:2007,traulsen:2010,grujic:2010}.
Arguably, the main conclusion of this research is that lattice-like structures
do not seem to promote cooperation, at least not to a extent different from what
is found in dyadic or small group experiments \citep{kagel:1995,camerer:2003}.
While the lack of promotion of cooperation 
is well established, the reasons proposed by the different teams to
explain the experimental observations are different, and there is no consensus
yet as to what is the way the subjects updated their decisions during the 
experiment. In particular, \cite{kirchkamp:2007} focused on disproving the
imitation strategy proposed by \cite{nowak:1992}, a conclusion also suppported
by \cite{grujic:2010}. On the other hand, \cite{traulsen:2010} fitted their
results to a payoff-dependent imitation behavior ---Fermi rule
\citep{szabo:1998}---, finding that
they needed a large amount of random mutation to explain their observations. 

In the above context, the analysis carried out by \cite{grujic:2010} brought
in an alternative way to understand the experimental observations by building
upon the idea of reciprocity \citep{trivers:1971}, i.e., the fact that
individuals behave depending on the actions of their partners in the past.
In iterated two-player games, this idea has been studied through the concept
of reactive strategies \citep{nowak:1989a,nowak:1989b,nowak:1990,nowak:1992bis}
(see \citep{sigmund:2010} for a comprehensive summary on this matter), the
most famous of which is Tit-For-Tat \citep{axelrod:1981}, given by playing
what the opponent played in the previous run. Reactive strategies generalize
this idea by considering that players choose their action among the available
ones with probabilities that depend on the previous action of the opponent. 
For the simple case of two strategies (say C and D), players choose C with
probability $p$ following a C from their partner and with probability $q$
after a D from their partner. Subsequently this idea was further developed
by considering memory-one reactive strategies \citep{nowak:1995,sigmund:2010},
in which the probabilities depend on the previous action of both the focal player
and her opponent ---i.e., the focal player would choose C with some probability
following a (C,C) outcome, and so on.   

\begin{figure*}
\centering
\includegraphics[width=0.45\textwidth,clip]{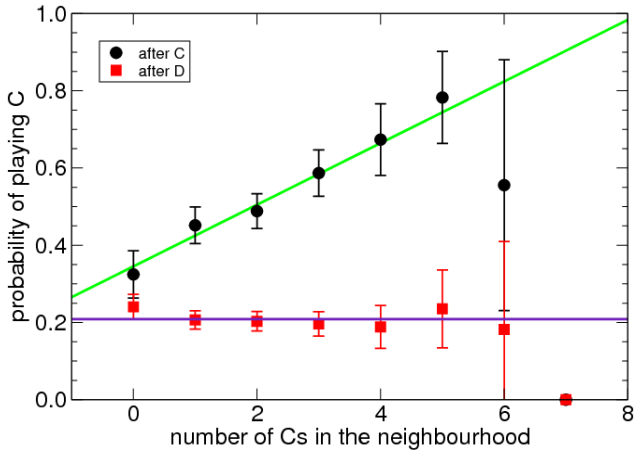}
\hspace*{5mm}
\includegraphics[width=0.45\textwidth,clip]{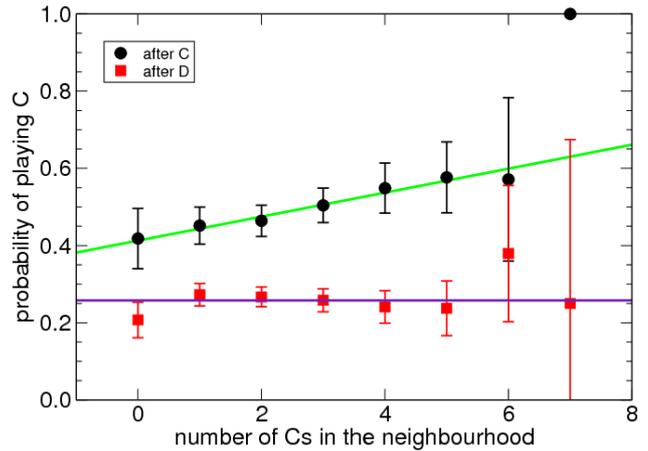}
\caption{Probabilities of cooperating after playing C or D, conditioned to
the context (number of cooperators in the previous round) in the two experiments
by \cite{grujic:2010}. Parameters of the fitted lines will be used
later as inputs for our replicator dynamics study. The line fitted to the
probabilities of cooperation after playing D is strictly horizontal.}
\label{fig:experiment}
\end{figure*}

In iterated multi-player games, such as public goods games or multi-player
Prisoner's Dilemmas (IMPD), reciprocity arises in the form of conditional
cooperation \citep{fischbacher:2001,gachter:2007}: individuals are willing
to contribute more to a public good the more others contribute. Conditional
cooperation has been observed a number of times in public goods experiments
\citep{croson:2006,fischbacher:2010}, often along with a large percentage
of free-riders. The experiment by \cite{traulsen:2010} showed also evidence
for such a behavior in an spatial setup. \cite{grujic:2010} extended this
idea in their analysis to include the dependence of the focal player's previous
action, introducing the so-called moody conditional cooperation
(cf.~Figure~\ref{fig:experiment}). In this strategy, players are
more prone to cooperate after having cooperated than after having defected,
and in the first case they are more cooperative the more cooperative neighbors
they have. This behavior has not been reported before in spatial games and
appears to be a natural extension of the reactive strategy idea to multi-player
games (among the very many other extensions one can conceive). On the other
hand, and from an economic viewpoint, which is an important part of the
analysis of human behavior, this type of strategy update scheme responds to
the often raised questions on payoff-based rules. In economic interactions
it is usually the case that agents perceive the others' actions but not how
much do they benefit from them, and therefore the use of action updates
depending, e.g., on the payoff differences, may be questionable. This
seems to be the case even if this information is explicitly supplied to
the players \citep{grujic:2010}.

Interestingly, the conclusion of \cite{grujic:2010} had a new feature as compared 
to the other two experiments \citep{kirchkamp:2007,traulsen:2010}, namely
the heterogeneity of the population: aside from the
already mentioned moody conditional cooperators, there were a large minority of
defectors, i.e., players that defected all or almost all the time, and a few cooperators,
that cooperated at practically all rounds. This heterogeneity, also found to be 
very important in public goods experiments \citep{fischbacher:2010} had also been
observed in four-player experiments by \cite{kurzban:2005}, who reported that
their subjects could be roughly classified in three main types, including defectors,
cooperators and conditional cooperators (called reciprocators in the original work),
albeit they did not check for dependences on the past actions of the players either. 
Both \cite{kurzban:2005} and \cite{grujic:2010} checked that the payoffs obtained 
by every type of player were more or less the same, thus suggesting that the 
population in the lattice experiment might be at an evolutionary equilibrium. 

In this paper we address the question of the 
existence and stability of such a heterogeneous or mixed equilibrium in the 
multiplayer iterated Prisoner's Dilemma. It is important to understand
that we are not addressing the issue of the evolutionary explanation of
moody conditional cooperation. This is a very interesting but also very
difficult task, and in fact we do not even have an intuition as to how
can one address this problem in a tractable manner. Our goal is then
to understand whether or not the coexistence of moody conditional
cooperators, defectors, and a small percentage of cooperators, as
observed in the experiment, is theoretically possible. In so doing,
we will shed light on experimental and theoretical issues at the same
time. On the experimental side, our results show that there is
coexistence for groups of 2 or 3 players for parameters
reasonably close to those found in the experiment, but not for larger
groups. As we will see in the discussion section, this prediction has
important consequences related to the adequacy of replicator dynamics
to describe the experimental result or to the cognitive capabilities
of human subjects in dealing with large groups. We will also discuss
there the ways in which our theoretical approach and the experiment
may differ, something that can also have implications of its own.
On the theoretical side, we present an analysis of a population of
players interacting through a multi-player Prisoner's Dilemma including
strategies that generalize the ideas behind reactive strategies, as
mentioned above. To our knowledge, this has not been carried out
before, at least to the extent we are doing it here, in which we
are able to show how this coexistence depends on the size of the
groups considered. We believe that the approach we are presenting
may be useful for other researchers working on related problems.

With the above goals in mind, we introduce below a model 
in which populations consisting of the three types of individuals discussed above,
namely cooperators, defectors, and moody conditional cooperators, play a 
multiplayer iterated Prisoner's Dilemma with populations evolving according to 
the replicator dynamics. We have considered different group sizes, from $n=2$ 
through $n=5$ players, a size whose outcome is well described by the limit
$n\to\infty$, which we analyze separately. In the 
following, we report on our findings concerning this system beginning by a 
detailed introduction of our model (Sec.\ 2). The key to our analytical approach
is the payoff matrix, whose computation we carry out by means of Markov chain
techniques. Section 3 presents the calculation in full detail for $n=2$ players
(i.e., the standard iterated Prisoner's Dilemma) and, subsequently,
proceeds to the replicator dynamics analysis of the so-obtained payoff matrix. 
We then extend our approach to larger groups ($n=3, 4, 5$) in Sec.\ 4. As this 
becomes more and more cumbersome, in Sec.\ 5 we address the
limit $n\to\infty$ analytically finding 
exact results about the possible equilibria of the model.  
Finally, Sec.\ 6 concludes by comparing our 
results to the experiments and discussing the implications of such a comparison 
in terms of theoretical explanations of the observed behavior and their 
shortcomings.

\section{Game, strategies and payoffs}

Let us consider a well-mixed population of players who interact via iterative
multiplayer prisoner's dilemmas (IMPDs). In these games, players interact in
groups of $n$ players. Every round each player adopts an action, either cooperate (C)
or defect (D), and receives a payoff from every other player in the group according
to a standard prisoner's dilemma payoff matrix (a cooperator receives $R$ from
another cooperator and $S$ from a defector; a defector receives $T$ from a
cooperator and $P$ from another defector; payoffs satisfy $T>R>P>S$). We note that
this is
a generalized version of a public goods game: In the latter, if there are $k$
cooperators, a defector receives $bk$ whereas a cooperator receives $b'(k-1)-c$
($b'=b$ in a standard public goods game). In an multiplayer PD, a defector receives
$(T-P)k+P(n-1)$ whereas a cooperator receives $(R-S)(k-1)+S(n-1)$, and hence
choosing 
$b=T-P$, $b'=R-S$ and $c=(P-S)(n-1)$ the IMPD becomes a generalized public
goods game. Notice an important difference with respect to the standard
public goods game: in this generalized version ($b\ne b'$) the difference
between the payoff received by a cooperator and a defector depends on the
number of cooperators.

Inspired by the experimental
results of \cite{grujic:2010} but keeping at the same time as few parameters as
possible, we will classify players' strategies into three stereotypical behaviors:
mostly cooperators, who cooperate with probability $p$ (assumed relatively close
to one) and defect with probability $1-p$; mostly defectors, who cooperate with
probability $1-p'$ and defect with probability $p'$ (for simplicity we will
assume $p'=p$); and moody conditional cooperators, who play depending on theirs
and their opponents' actions in the previous round. Specifically, if they defected
in the previous round they will cooperate with probability $q$, whereas if
they cooperated in the previous round they will cooperate again with a probability
\begin{equation}
p_C(x)=(1-x)p_0+xp_1
\label{eq:condcoopprob}
\end{equation}
where $x$ is the fraction of cooperative actions
among the opponents in the previous round, and $p_0<p_1$. 

At 
this point, we would like to mention 
that our results do not depend qualitatively on the ``moodiness assumption''; in fact, we 
have checked that redoing the calculations we will present below for plain conditional cooperators (as 
those found by \cite{fischbacher:2001,gachter:2007}) leads only to quantitative changes in the 
results. Therefore, we present our discussion in terms of moody conditional cooperators as they 
are empirically more relevant. 

To complete the definition of the model, we 
need to specify how the populations of the different strategies 
are going to evolve in time. 
Players interact infinitely often in an IMPD, so payoffs both increase
in time and depend on the whole history of play. It thus make sense to
use the (time) average payoffs to study the evolution of the game in terms
of the abundance of the three strategies considered. As
these strategies are defined depending on players' actions in the
round immediately before, a multiplayer game with $n$ players and
given populations of each type of
player can be described as a finite state Markov chain
whose states are defined by the actions taken by
the $n$ players. Of course the chain is different
for different compositions of strategies in the group. In any case, given
that all outcomes have non-zero probability, the chain is ergodic and
therefore there is a well defined steady state \citep{karlin:1975}.
Average payoffs are readily obtained once the probability vector in the
steady state is known, and subsequent evolution is described through imitation via
replicator dynamics \citep{hofbauer:1998}. In the next section we develop all this
formalism in full detail for the case $n=2$, i.e., for the usual iterated PD, taking 
advantage that in this case the expressions that arise can be written in a compact
way. The cases with more than 2 players are dealt with in Sec.\ 4 in a more 
sketchy manner. 

\section{Two-persons game (iterated PD)}

\subsection{General scheme of the approach}

In the case $n=2$ the states of the Markov chain are described as
CC, CD, DC, and DD, where the first action is the focal player's and
the second one is the opponent's. The transition probability matrix
will be denoted as
\begin{equation}
M=
\bordermatrix{
     & $CC$   & $CD$   & $DC$   & $DD$   \cr
$CC$ & m_{11} & m_{12} & m_{13} & m_{14} \cr
$CD$ & m_{21} & m_{22} & m_{23} & m_{24} \cr
$DC$ & m_{31} & m_{32} & m_{33} & m_{34} \cr
$DD$ & m_{41} & m_{42} & m_{43} & m_{44} \cr
},
\label{eq:Markov}
\end{equation}
where $m_{ij}$ gives the probability that players who played $i$ in
the previous round play $j$ in the present round
($i,j\in\{$CC, CD, DC, DD$\}$). The matrix $M$ will of course depend on the
nature of the two players involved, so there will be nine different
matrices. Denoting `mostly cooperators' by C, `mostly defectors' by D
and `moody conditional cooperators' by X, the six combinations are CC, CD, CX,
DD, DX, XX. As we stated above, the  Markov chains so defined are always 
ergodic; consequently, the corresponding stationary probability vector, which 
we will term
$\pi=(\pi_{\rm CC},\pi_{\rm CD},\pi_{\rm DC},\pi_{\rm DD})$, 
is obtained by solving the equation $\pi=\pi M$ \citep{karlin:1975}.
Note that there is such a stationary probability distribution $\pi$ for each of the six
combinations of two players, as we will see below. Now, 
once the probability distribution is known, 
the payoff matrix $W=(w_{ij})$, providing the average payoff that
a player of type $i$ gets when confronted to a player of type $j$
($i,j\in\{$C, D, X$\}$) in an IMPD (in this Section, $n=2$, an iterated PD) 
can be computed as
\begin{equation}
w_{ij}=R\pi_{\rm CC}+S\pi_{\rm CD}+T\pi_{\rm DC}+P\pi_{\rm DD}.
\label{etiq2}
\end{equation}
These payoffs can then be used in the replicator dynamics to finally find
the evolution of the three strategy population. 

\subsection{Payoff computation}

Of the six combinations of players, three yield a trivial stationary
vector $\pi$  because
they do not depend on the previous actions, namely those which do not involve the strategy X. 
The corresponding payoffs are therefore 
straightforward to compute, and we have (recall the focal player is denoted by the first 
subindex): 
\begin{align}
w_{\text{CC}} &= p^2 R + p(1-p) S + (1-p)p T + (1-p)^2 P, \\
w_{\text{CD}} &= p(1-p) R + p^2 S + (1-p)^2 T + (1-p)p P, \\
w_{\text{DC}} &= (1-p)p R + (1-p)^2 S + p^2 T + p(1-p) P, \\
w_{\text{DD}} &= (1-p)^2 R + (1-p)p S + p(1-p) T + p^2 P.
\end{align}

The payoffs for the cases where the moody conditional cooperators, X, play, require 
computing the corresponding stationary probability. Let us begin with the 
Markov matrix \eqref{eq:Markov} for a mostly cooperator (C) and a
conditional cooperator (X), given by 
\begin{equation} 
M=
\begin{pmatrix}
 p p_1 & p \left(1-p_1\right) & (1-p)
   p_1 & (1-p) \left(1-p_1\right) \\
 p q & p \left(1-q\right) & (1-p) q & (1-p)
   \left(1-q\right) \\
 p p_0 & p \left(1-p_0\right) & (1-p)
   p_0 & (1-p) \left(1-p_0\right) \\
 p q & p \left(1-q\right) & (1-p) q & (1-p)
   \left(1-q\right)
\end{pmatrix},
\end{equation}
from which the stationary probability vector is given by
\begin{equation}
\pi =\frac{\big(pq, p[1-p_C(p)], (1-p)q, (1-p)[1-p_C(p)]\big)}{1+q-p_C(p)},
\label{etiq1}
\end{equation}
where $p_C(x)$ is given by \eqref{eq:condcoopprob} (notice that it
represents the average probability for a conditional cooperator to
cooperate, given that she cooperated in the previous round, whereas
her mostly cooperator 
opponent cooperates with probability $p$). Therefore, inserting (\ref{etiq1}) in
(\ref{etiq2}) and having in mind who the focal player is, we arrive at 
\begin{align}
 w_{\text{CX}}=&[1+q-p_C(p)]^{-1}\Big\{pqR+p[1-p_C(p)]S+(1-p)qT \nonumber\\
               &+(1-p)[1-p_C(p)]P\Big\}\\
 w_{\text{XC}}=&[1+q-p_C(p)]^{-1}\Big\{pqR+(1-p)qS+p[1-p_C(p)]T \nonumber\\
               &+(1-p)[1-p_C(p)]P\Big\}.
\end{align}

The case for a
mostly defector facing a moody conditional cooperator can be obtained 
immediately by realizing that the defector behaves as a mostly cooperator whose probability
of cooperating is $1-p$ instead of $p$, hence we find trivially
\begin{align}
 w_{\text{DX}}=&[1+q-p_C(1-p)]^{-1}\Big\{(1-p)qR + pqT \nonumber\\
               &+(1-p)[1-p_C(1-p)]S+p[1-p_C(1-p)]P\Big\}, \\
 w_{\text{XD}}=&[1+q-p_C(1-p)]^{-1}\Big\{(1-p)qR+pqS \nonumber\\
               &+(1-p)[1-p_C(1-p)]T+p[1-p_C(1-p)]P\Big\}.
\end{align}

Finally, if two conditional cooperators confront each other, the Markov
matrix becomes
\begin{equation}
M=
\begin{pmatrix}
 p_1^2 & p_1\left(1-p_1\right) &
   \left(1-p_1\right) p_1 &
   \left(1-p_1\right){}^2 \\
 p_0 q & p_0 \left(1-q\right) &
   \left(1-p_0\right) q &
   \left(1-p_0\right) \left(1-q\right) \\
 q p_0 & q \left(1-p_0\right) &
   \left(1-q\right) p_0 &
   \left(1-q\right) \left(1-p_0\right) \\
 q^2 & q \left(1-q\right) & \left(1-q\right) q &
   \left(1-q\right){}^2
\end{pmatrix},
\end{equation}
and has a stationary vector $\pi$ which, up to normalization, is proportional
to a vector $\alpha$ with components
\begin{equation}
\begin{split}
\alpha_{\text{CC}} &=q^2(1+p_0-q), \\
\alpha_{\text{CD}} &=q(1-p_1)(1+p_1-q),\\
\alpha_{\text{DC}} &=q(1-p_1)(1+p_1-q),\\
\alpha_{\text{DD}} &=(1-p_1^2)(1-p_0-q)+2qp_0(1-p_1).
\end{split}
\end{equation}
From this result one can compute $w_{\text{XX}}$ as in the other eight cases.
With the payoffs we have computed, we are now in a position to proceed to
the dynamical study. 

\subsection{Replicator dynamics}

Denoting $x=(x_{\text{C}},x_{\text{D}},x_{\text{X}})$ (with
$x_{\text{C}}+x_{\text{D}}+x_{\text{X}}=1$) the vector with the population
fractions of the three types of players, the dynamics of $x_i$ is described
by the replicator equation
\begin{equation}
 \dot{x_i}=x_i[ (W x)_i-x\cdot W x ],
\label{repdyn}
\end{equation}
where $W$ is the payoff matrix obtained above. 

\begin{figure}[t]
\centering
\includegraphics[width=0.30\textwidth,clip]{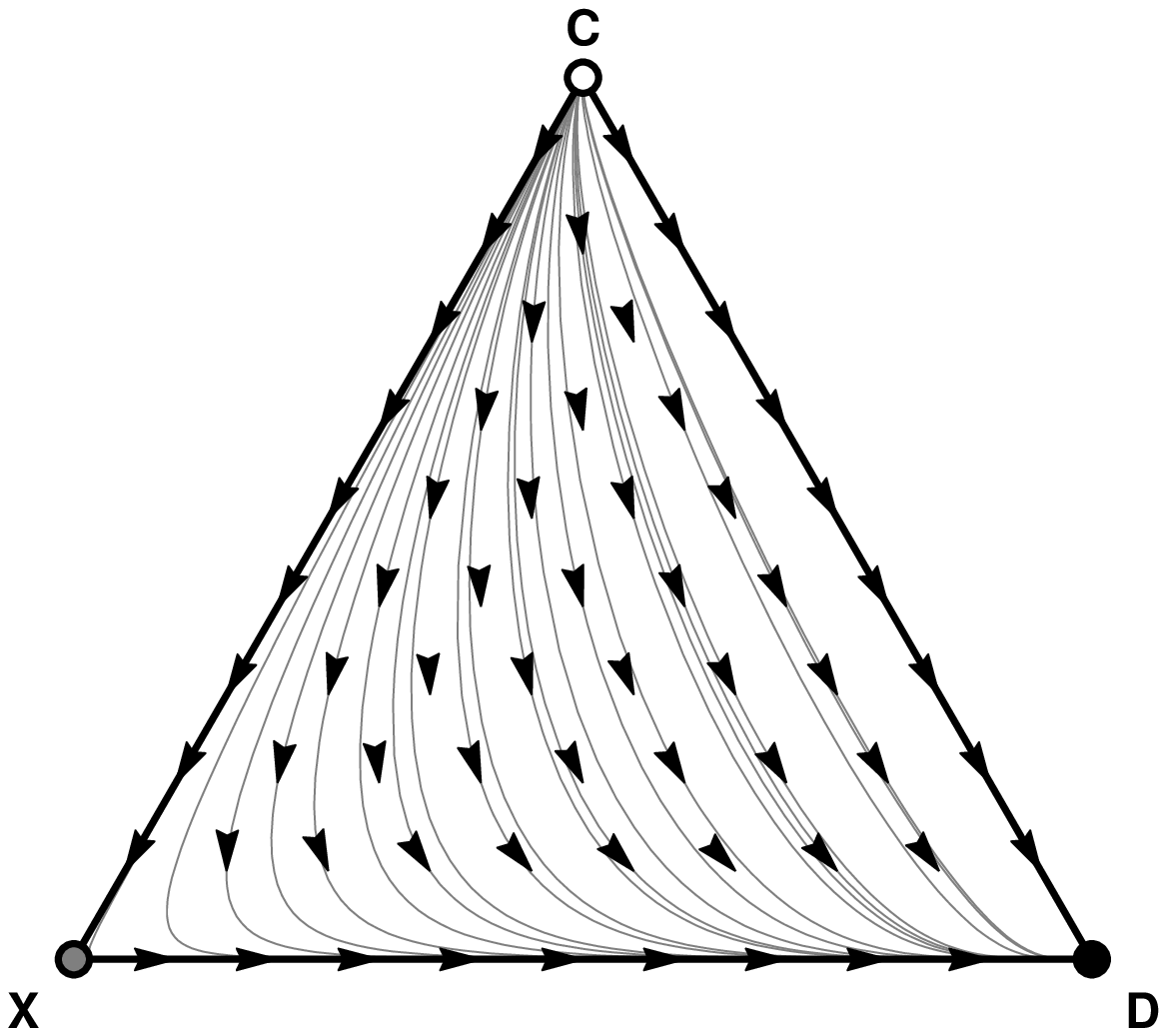}\\
\includegraphics[width=0.30\textwidth,clip]{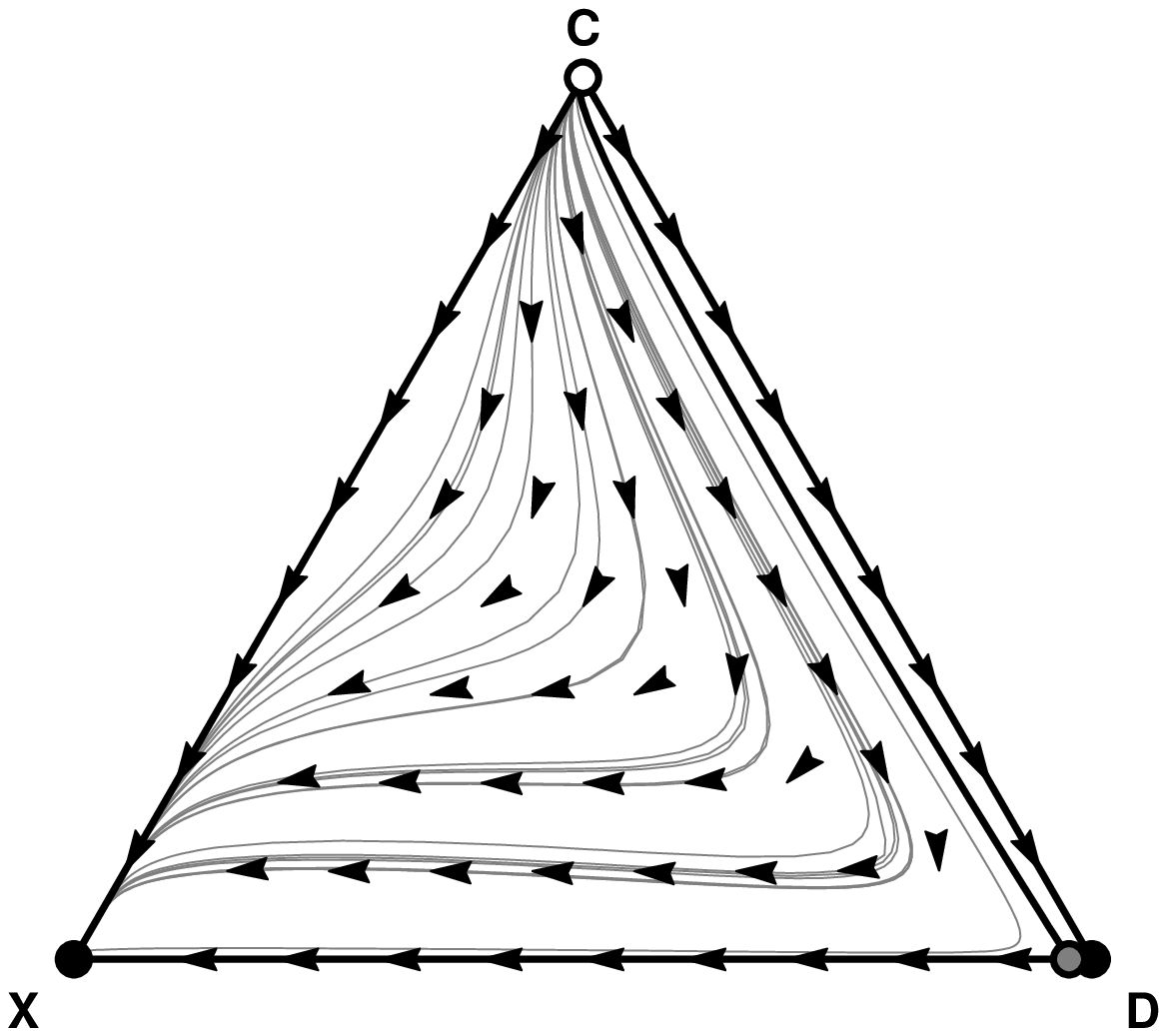}
\caption{Phase portraits of the replicator dynamics for 2-players IMPD games
with three strategies (C, D, and X) for the parameters inferred from
experiment 1 (top; $p=0.83$, $q=0.26$, $p_0=0.44$, $p_1=0.60$) and
experiment 2 (bottom; $p=0.83$, $q=0.21$, $p_0=0.34$, $p_1=0.98$).
Rest points marked in the plot 
can be repellors (white), saddle points (grey) or  attractors (black).}
\label{fig:experimentalpars}
\end{figure}

In order to use this dynamics in connection with the experiment of \cite{grujic:2010},
we need to recall the payoffs used in that work, namely 
$T=10$, $R=7$, $P=S=0$ [i.e., a weak prisoner's
dilemma as in \cite{nowak:1992}]. Two consecutive experiments were 
carried out, leading to two different sets of parameters for the behavior
of the players. Figure~\ref{fig:experimentalpars} shows the dynamics 
resulting for both sets of parameters, whose specific values are listed in
the caption. As we may see, there are no interior points, which would 
indicate equilibria in which the three strategies coexist, as observed
in the experiment. The only equilibria we find for these parameters are 
in the corners of the simplex, C being always a repeller, D an attractor
and X being a saddle point or an attractor depending on the parameters.
In the case where D and X are both attractors it is X that has the
largest basin of attraction (almost the entire simplex),
Therefore, the results for
this model do not match what is observed in the experiment. However, 
it is important to keep in mind that in the experiment players played with
their eight neighbors, this being the reason why we will later address 
the dynamics of IMPDs with larger groups. 

Notwithstanding this first result, as we will now see it is very interesting
to dwell into the $n=2$ case in more detail. For the purpose of illustrating
our results, let us choose the behavioral parameters to be  
$p=0.83$, $q=0.20$,
$p_0=0.40$, and $p_1=0.80$, which are values we could consider 
representative of both experiments.  
Inserting these parameters into the calculations above, we find that 
the payoff matrix is given by 
\begin{equation}
\begin{pmatrix}
 0      & -0.3366 &  \phantom{-}0.4367 \\
 1.6434 & 0       & -0.1800 \\
 1.0026 & -0.0526 & 0
\end{pmatrix}.
\label{eq:matrix083020040080}
\end{equation}
This type of matrix belongs to a class of games studied by \cite{zeeman:1980} 
[compare it with matrix \eqref{eq:Zeemanpayoff} in  \ref{sec:appendix}].
In fact, in a region of parameters near those that can be inferred from
the experiments of \cite{grujic:2010} the game behaves as a Zeeman game. 
The Zeeman game has five rest points (see \ref{sec:appendix}): an unstable one
at the C corner, a stable one at the D corner, a saddle point at the X corner,
and two mixed equilibria on the C--X and on the D--X edges of the simplex.
Besides, under certain constraints (c.f.~\eqref{eq:interior}) there is also
an interior point. 

\begin{figure*}
\centering
\includegraphics[width=0.30\textwidth,clip]{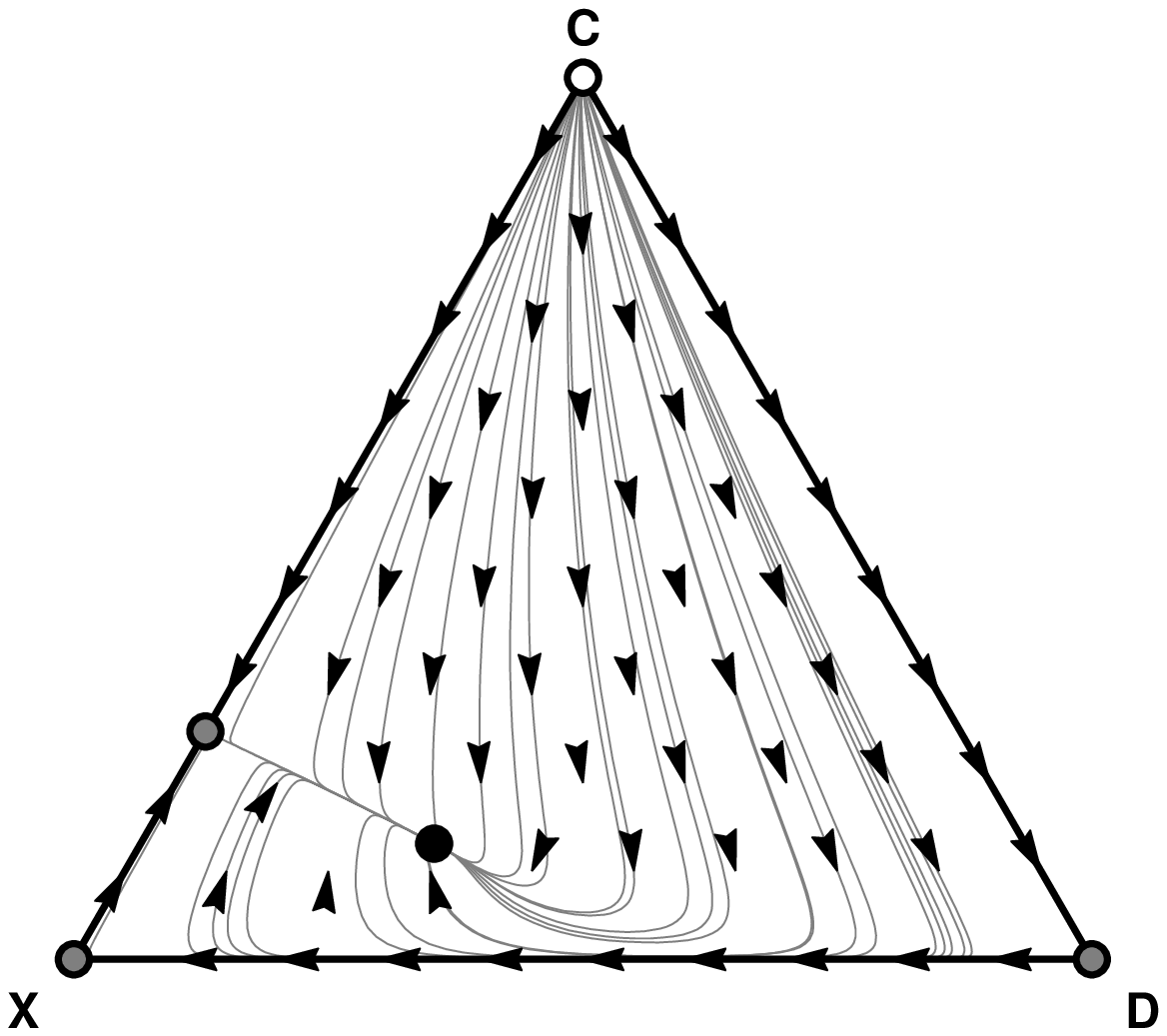}\hspace*{5mm}
\includegraphics[width=0.30\textwidth,clip]{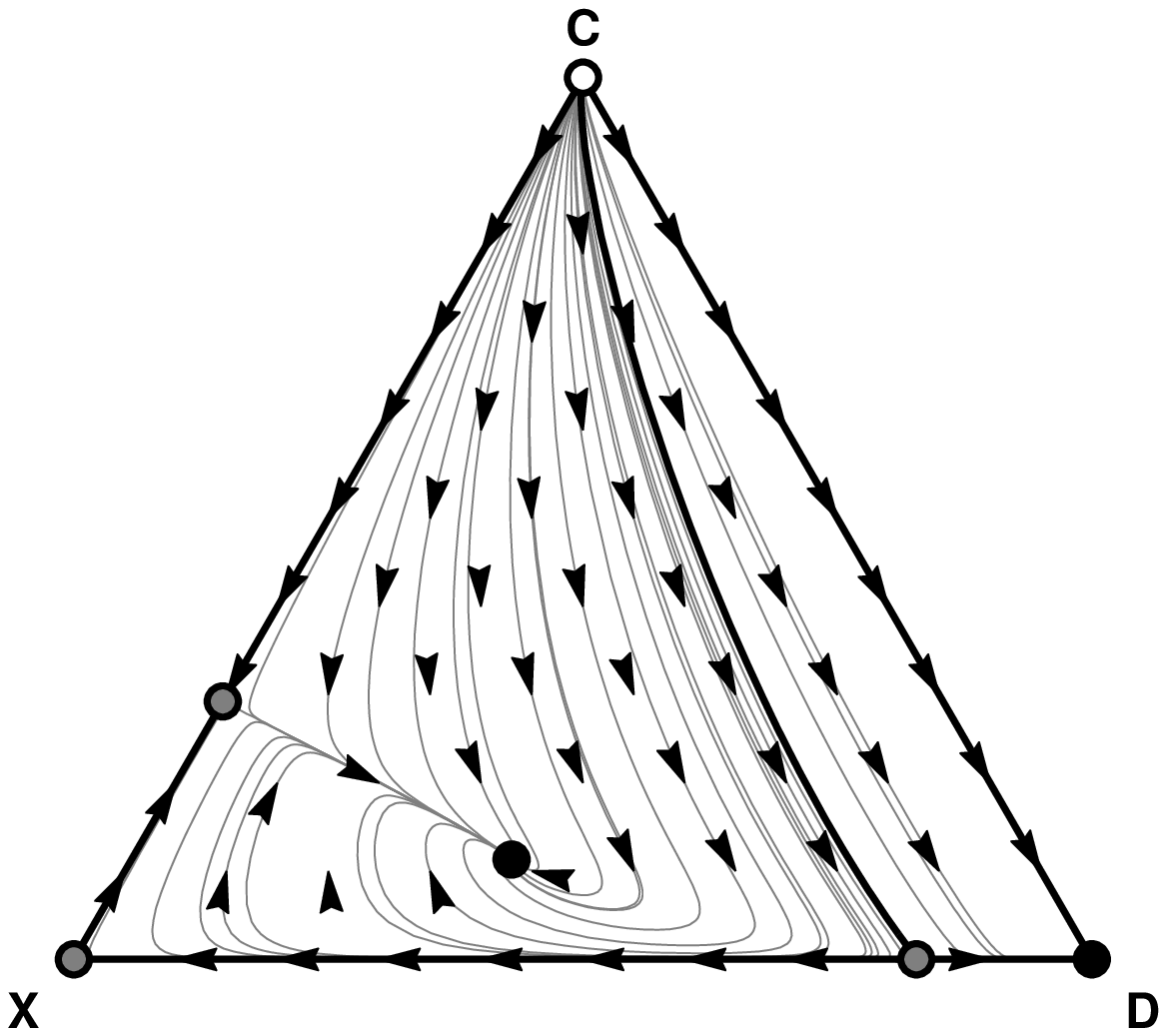}\hspace*{5mm}
\includegraphics[width=0.30\textwidth,clip]{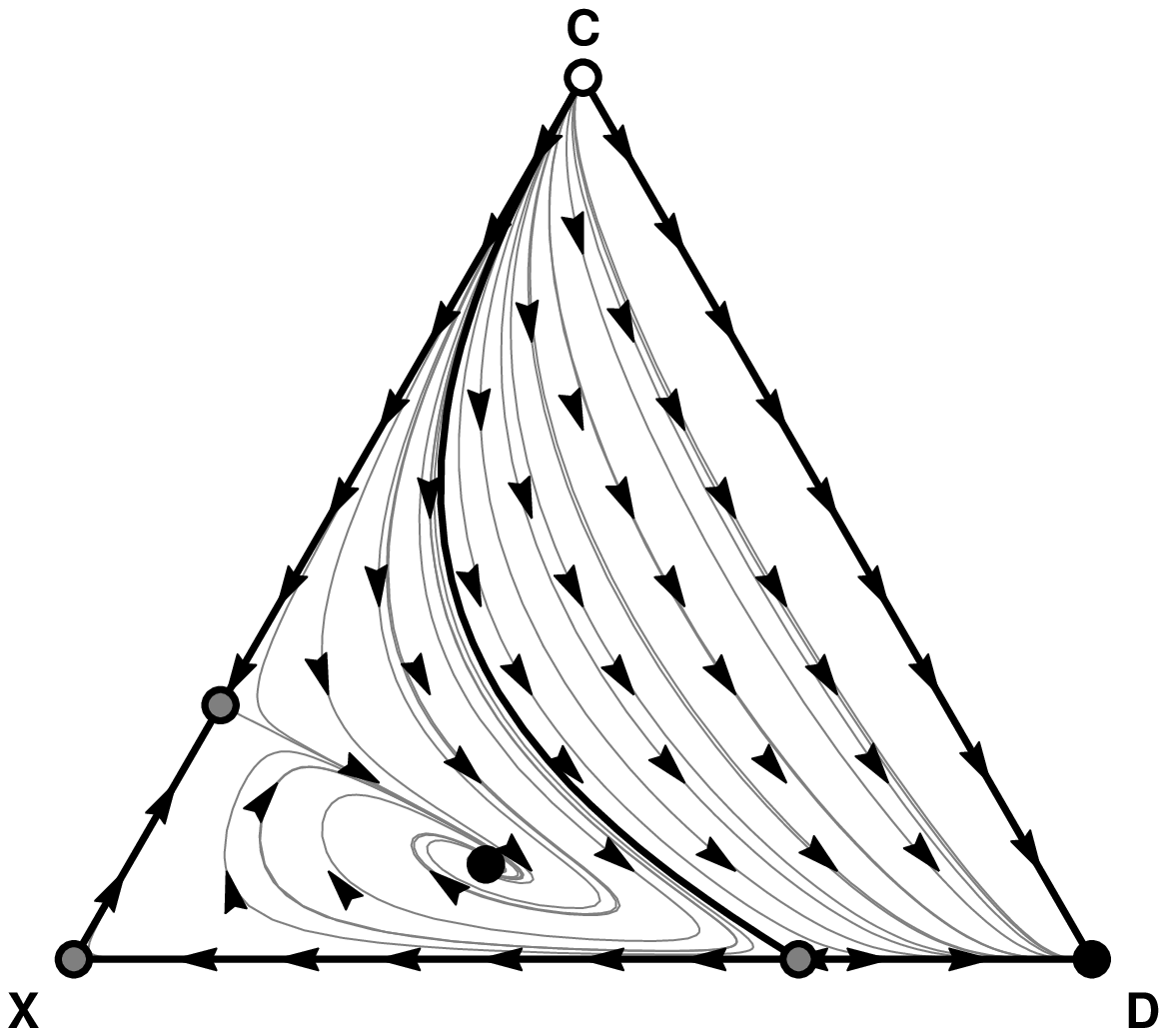}
\caption{Phase portraits of the replicator dynamics for 2-players IMPD games
with three strategies (C, D, and X) for different values
of $q$: $0.10$ (left), $0.15$ (middle), and $0.30$ (right). Other parameters:
$p=0.83$, $p_0=0.4$, $p_1=0.8$. Rest points marked in the plot 
can be repellors (white), saddle points (grey) or  attractors (black). In
the last two cases the inner point as well as the $x_D=1$ point are the only
attractors of the system. In the latter case the point $x_D=1$ has merged
with the saddle in the edge $x_C=0$ becoming a saddle point. Correspondingly,
the basin of attraction of $x_D=1$ has disappeared.}
\label{fig:2p-diff_q}
\end{figure*}

\begin{figure*}
\centering
\includegraphics[width=0.30\textwidth,clip]{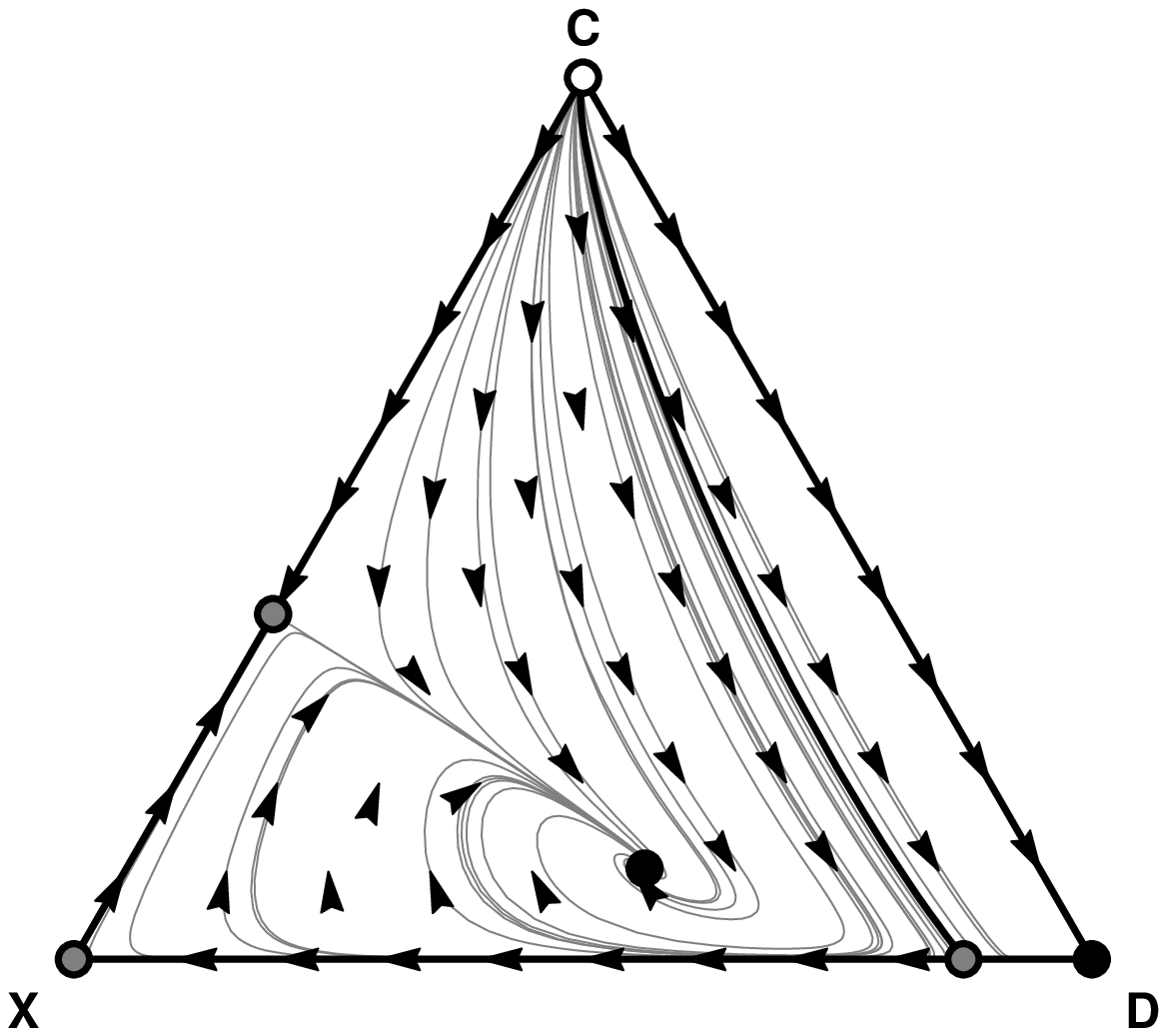}\hspace*{5mm}
\includegraphics[width=0.30\textwidth,clip]{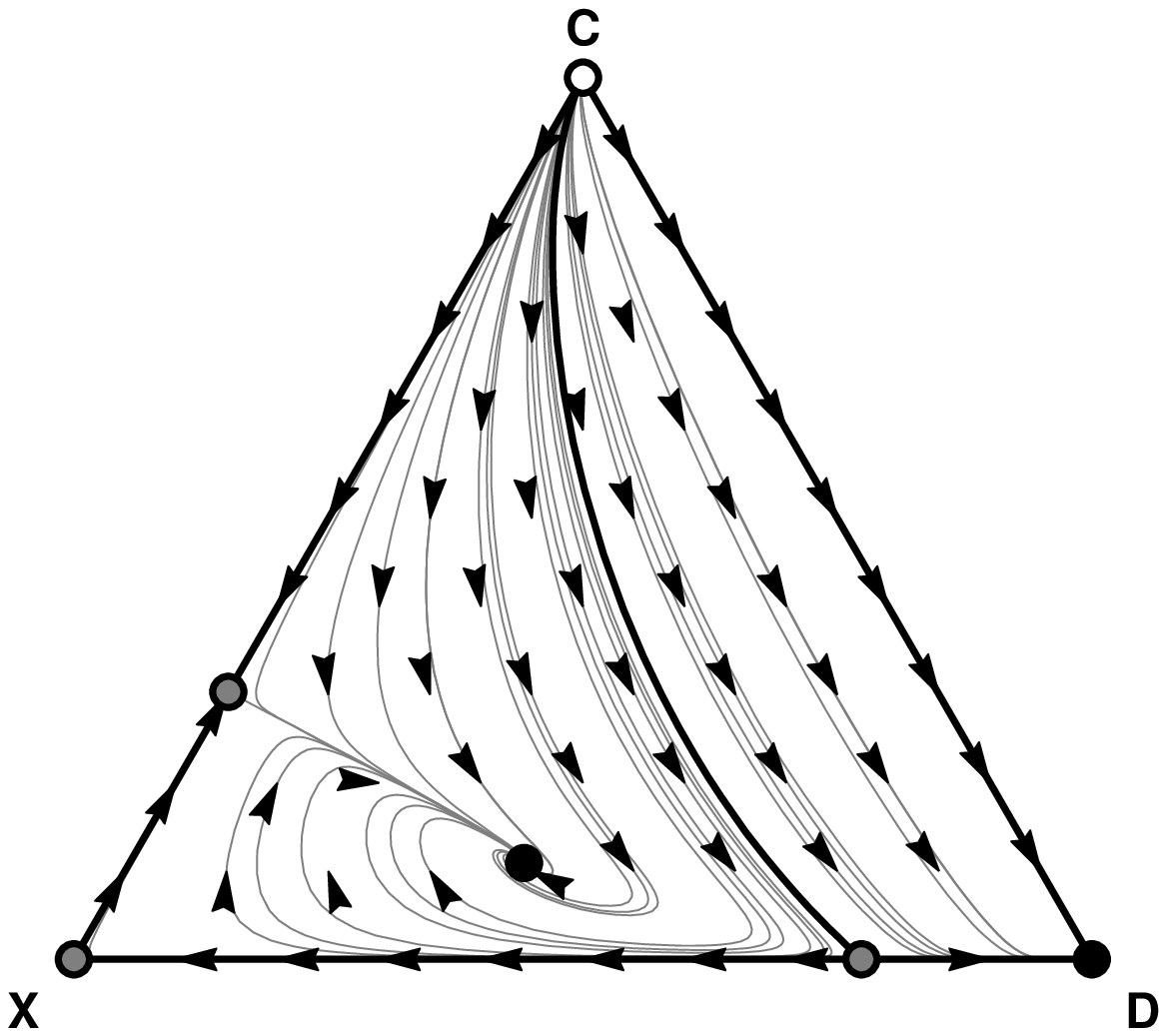}\hspace*{5mm}
\includegraphics[width=0.30\textwidth,clip]{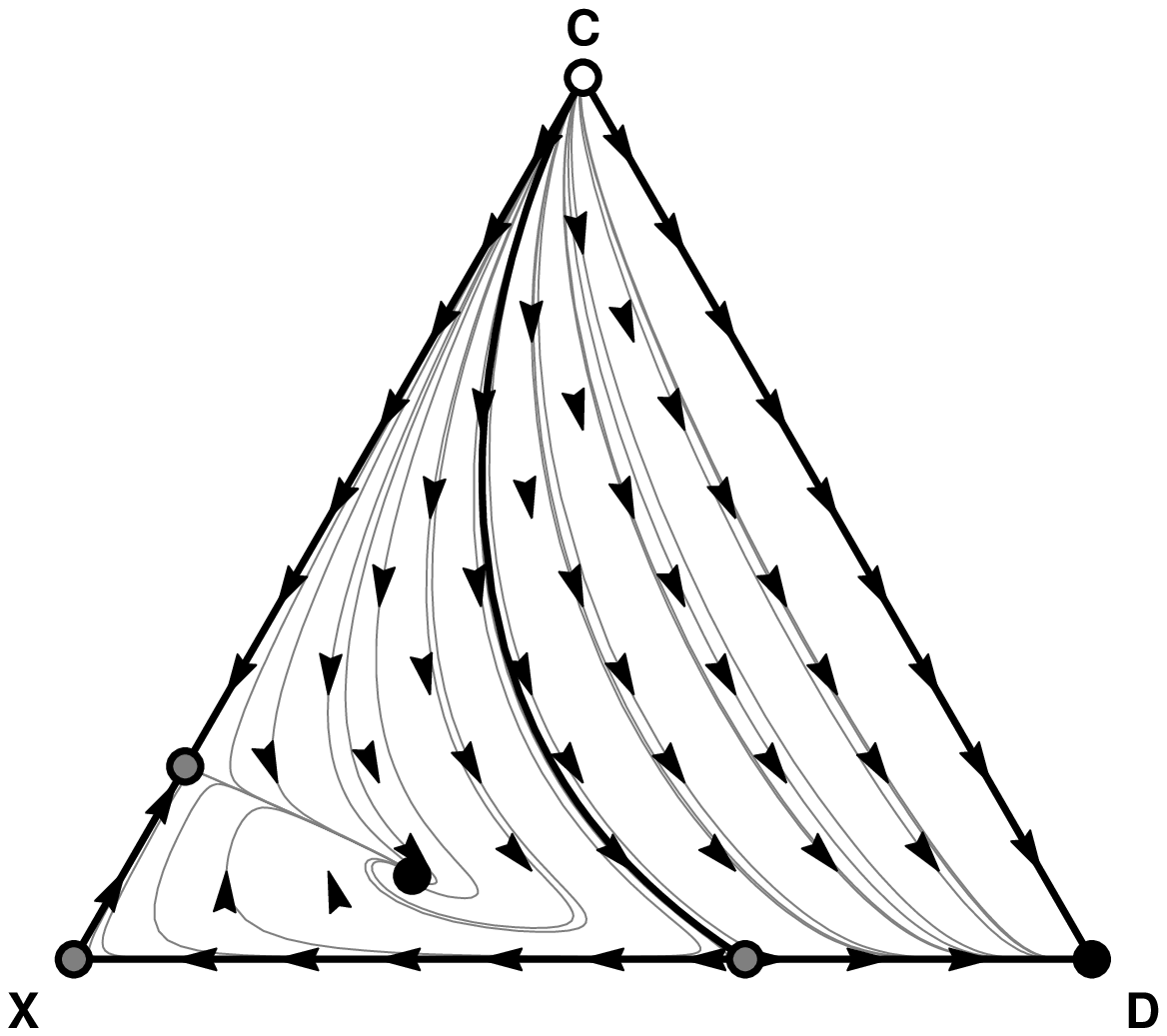}
\caption{Phase portraits of the replicator dynamics for 2-players IMPD games
with three strategies (C, D, and X) for different values
of $p_0$: $0.20$ (left), $0.40$ (middle), and $0.50$ (right). Other parameters:
$p=0.83$, $q=0.2$, $p_1=0.8$. Rest points marked in the plot 
can be repellors (white), saddle points (grey) or  attractors (black). In
all three cases the inner point as well as the $x_D=1$ point are the only
attractors of the system.}
\label{fig:2p-diff_p0}
\end{figure*}

\begin{figure*}
\centering
\includegraphics[width=0.30\textwidth,clip]{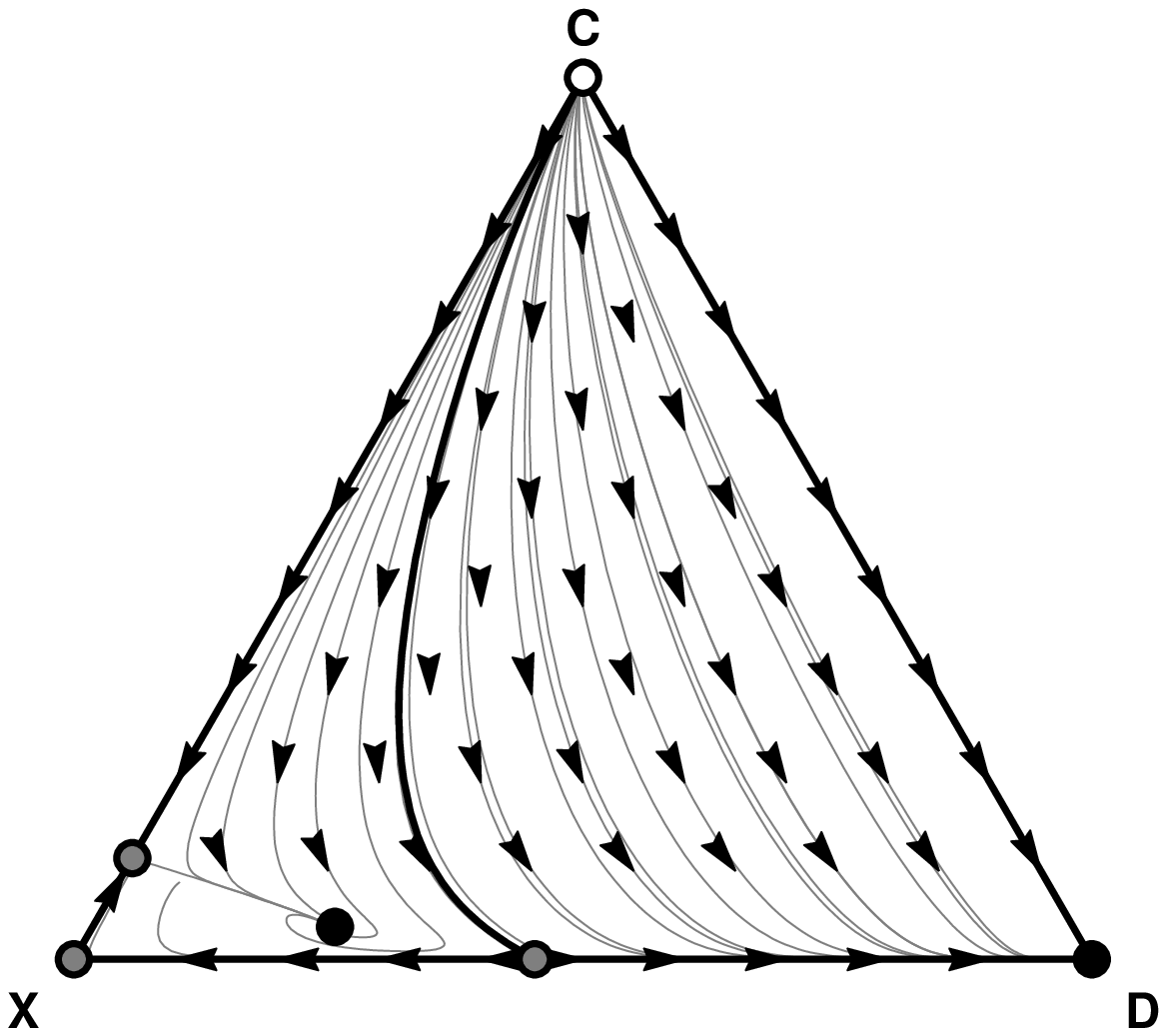}\hspace*{5mm}
\includegraphics[width=0.30\textwidth,clip]{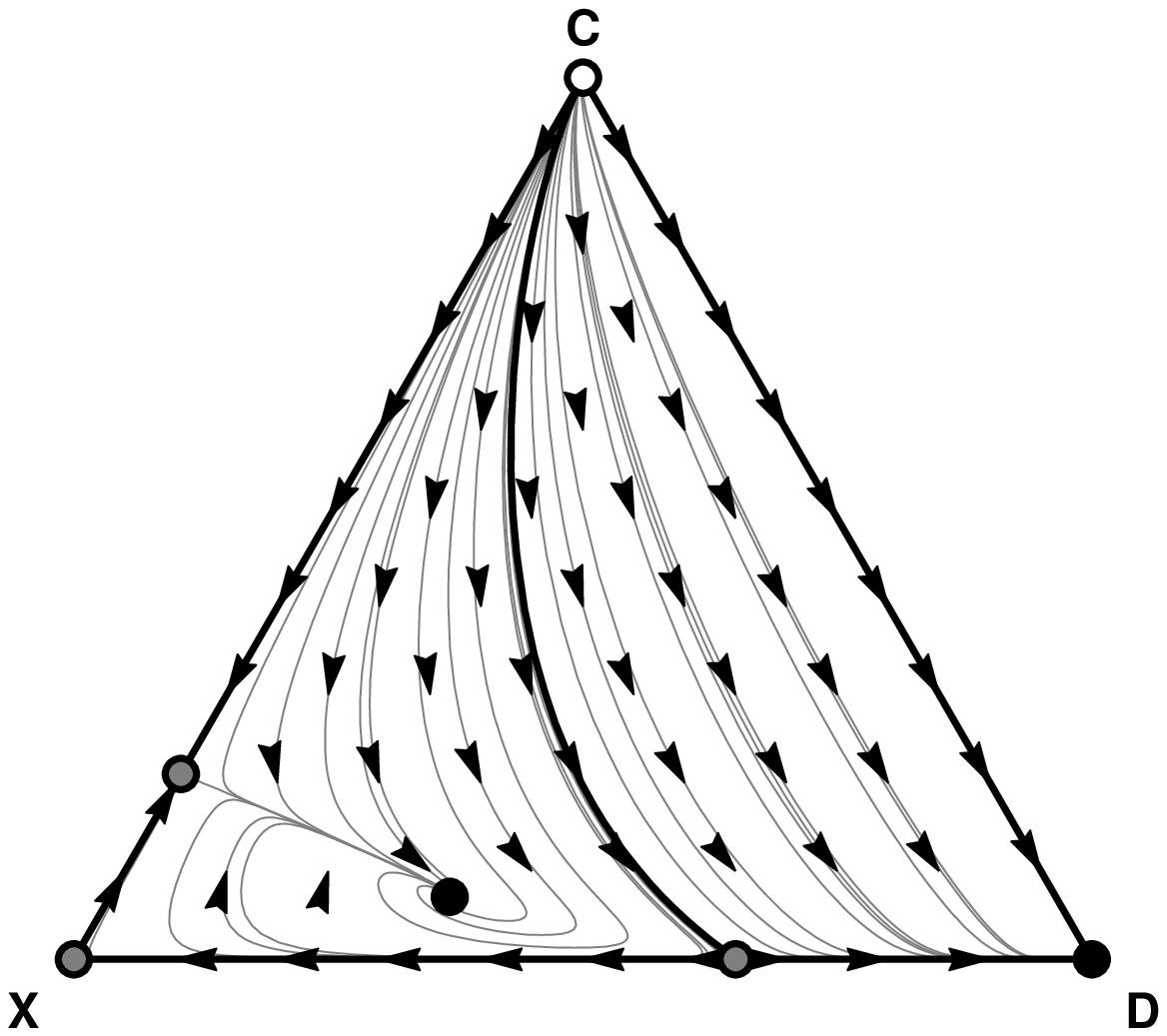}\hspace*{5mm}
\includegraphics[width=0.30\textwidth,clip]{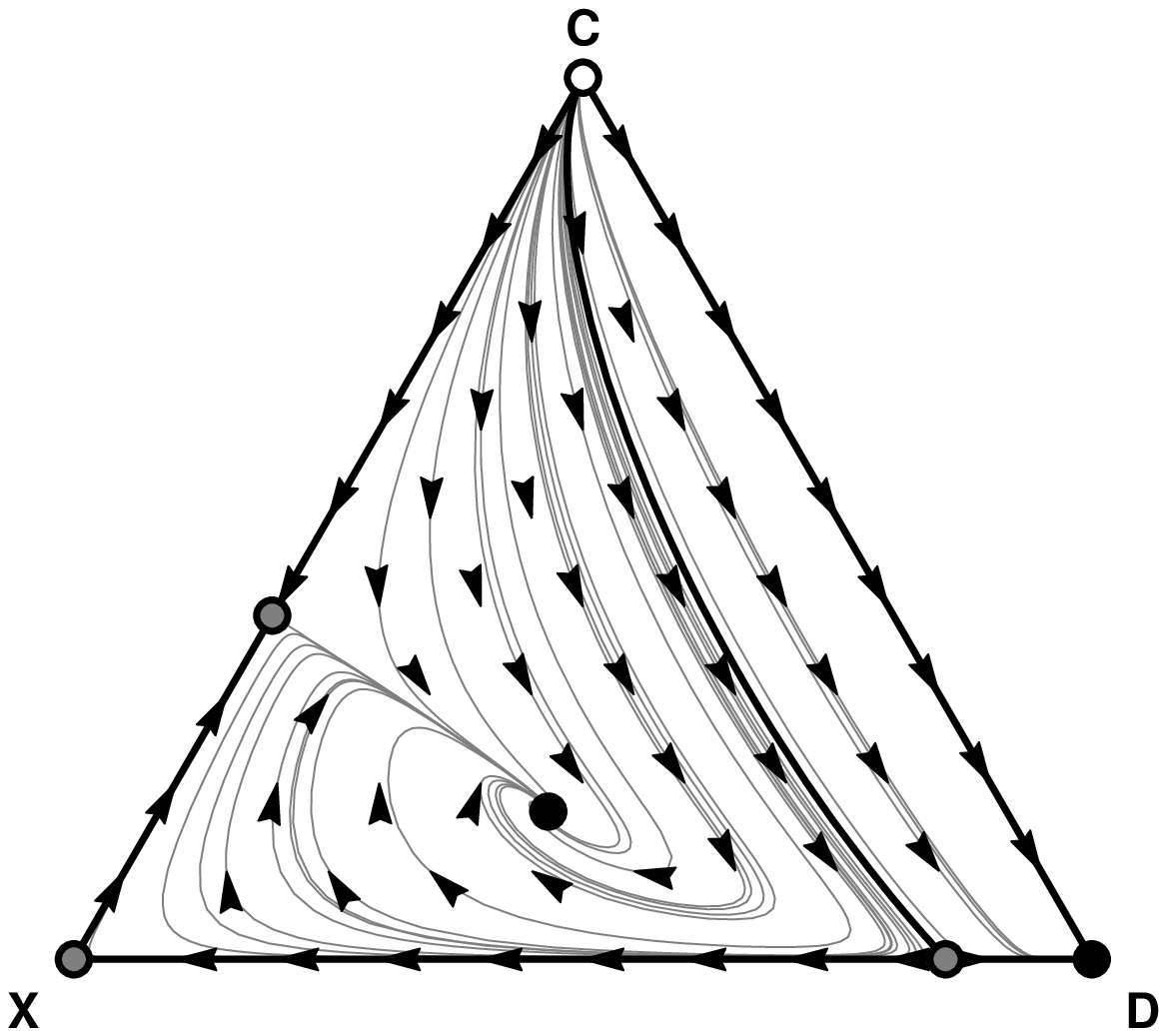}
\caption{Phase portraits of the replicator dynamics for 2-players IMPD games
with three strategies (C, D, and X) for different values
of $p_1$: $0.70$ (left), $0.75$ (middle), and $0.85$ (right). Other parameters:
$p=0.83$, $q=0.2$, $p_0=0.4$. Rest points marked in the plot 
can be repellors (white), saddle points (grey) or  attractors (black). In
all three cases the inner point as well as the $x_D=1$ point are the only
attractors of the system.}
\label{fig:2p-diff_p1}
\end{figure*}

Turning now to our example matrix \eqref{eq:matrix083020040080}, its non-trivial
rest points turn out to be $(0, 0.7739, 0.2261)$, $(0.3034, 0, 0.6966)$, and
$(0.1093, 0.3876, 0.5031)$. The stability of these mixed, interior equilibria depends
on the parameters (see \ref{sec:appendix}). For the present case, the situation
is similar to that shown in Figure~\ref{fig:zeeman}(a). Thus the evolution
of this system is governed by the presence of two attractors: the interior
point and the D corner, each with a certain basin of attraction. A key feature 
of the class of problems we are considering is that the 
precise location of the interior rest point is very sensitive to the values
of the parameters. Figures~\ref{fig:2p-diff_p}--\ref{fig:2p-diff_p1} illustrate
what happens to it when each of the four probabilities that define the strategies
are changed around the values given above.
Generally speaking, the figures show that the interior point approaches either
one of the rest points on the edges C--X and D---X, while these in turn move along
their edges. The specific details depend on the parameter one is considering as
can be seen from the plots.
We have also found that larger changes in the parameters
can make the interior point coalesce with the mixed equilibrium on the C--X
edge ---thus transforming the dynamics into the one sketched in
Figure~\ref{fig:zeeman}(c)--- or even change the Zeeman structure of the
payoff matrix yielding different stable equilibria (generally at the corners).
Notice that ---particularly so in experiment 2--- the values of the parameters
are not far from those producing the plots of
Figures~\ref{fig:2p-diff_p}--\ref{fig:2p-diff_p1}. This indicates that,
while we would
not expect a two-person theory to describe quantitatively the experiments, the 
existence of an interior point with the same kind of mixed population as observed 
is possible with minor modifications of the parameters.

\section{Games involving more than two players}

Having discussed in depth the replicator dynamics for the IPD with mostly
cooperators, mostly defectors and moody conditional cooperators, with the
result that an interior point with a sizable basin of attraction exists for a wide
range of parameters, we now increase the number of players to check 
whether the theory is a valid description of the experimental results. 
The mathematical approach for the case 
when more that two players are involved is similar to that
for two players, only computationally more involved. The Markov transition
matrix \eqref{eq:Markov} now describes a chain containing $2^n$ states,
$n$ being the number of players. These are described as all combinations of
C or D actions adopted by each of the $n$ interacting agents. On the other
\begin{figure*}
\centering
\includegraphics[width=0.30\textwidth,clip]{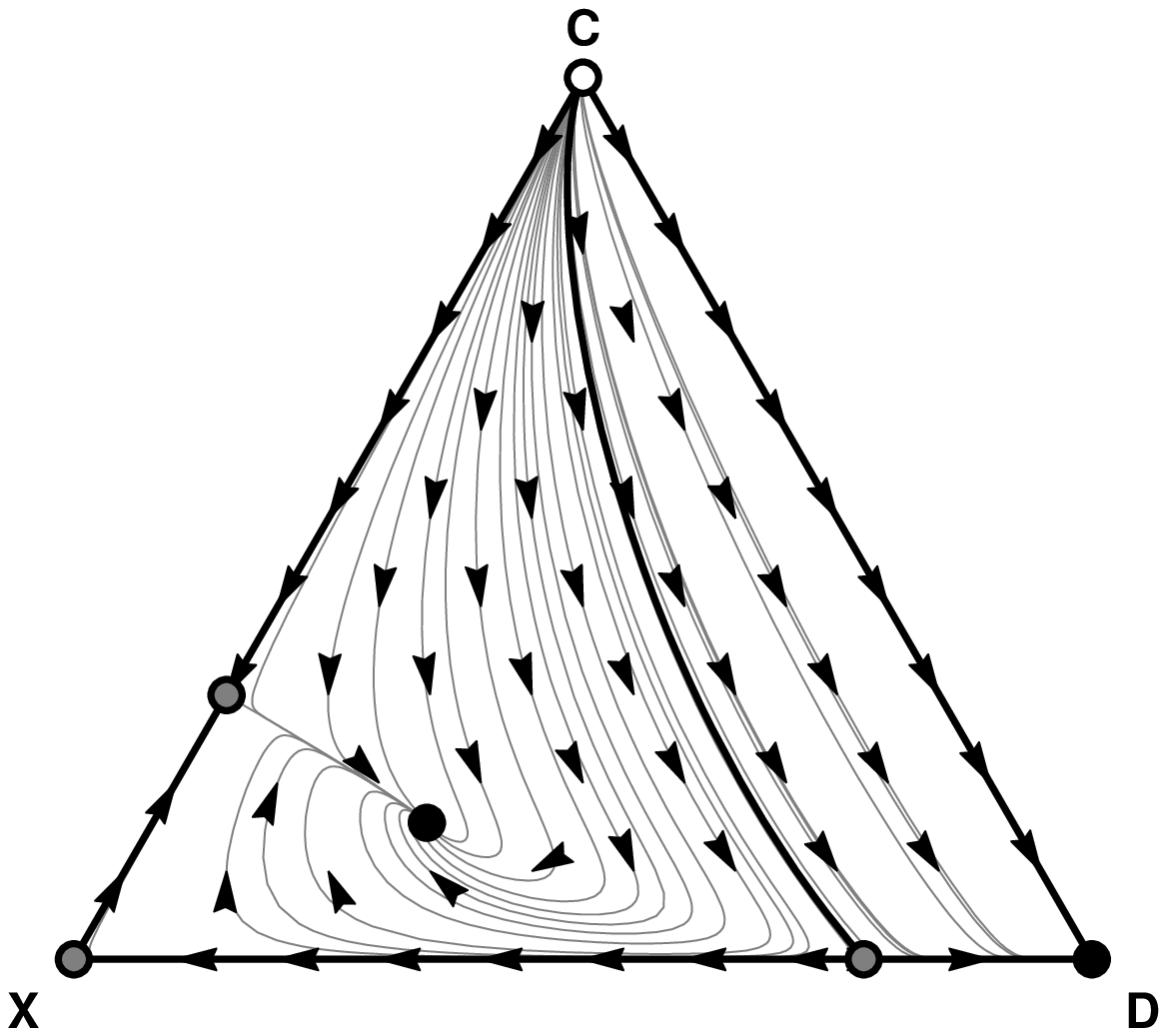}\hspace*{5mm}
\includegraphics[width=0.30\textwidth,clip]{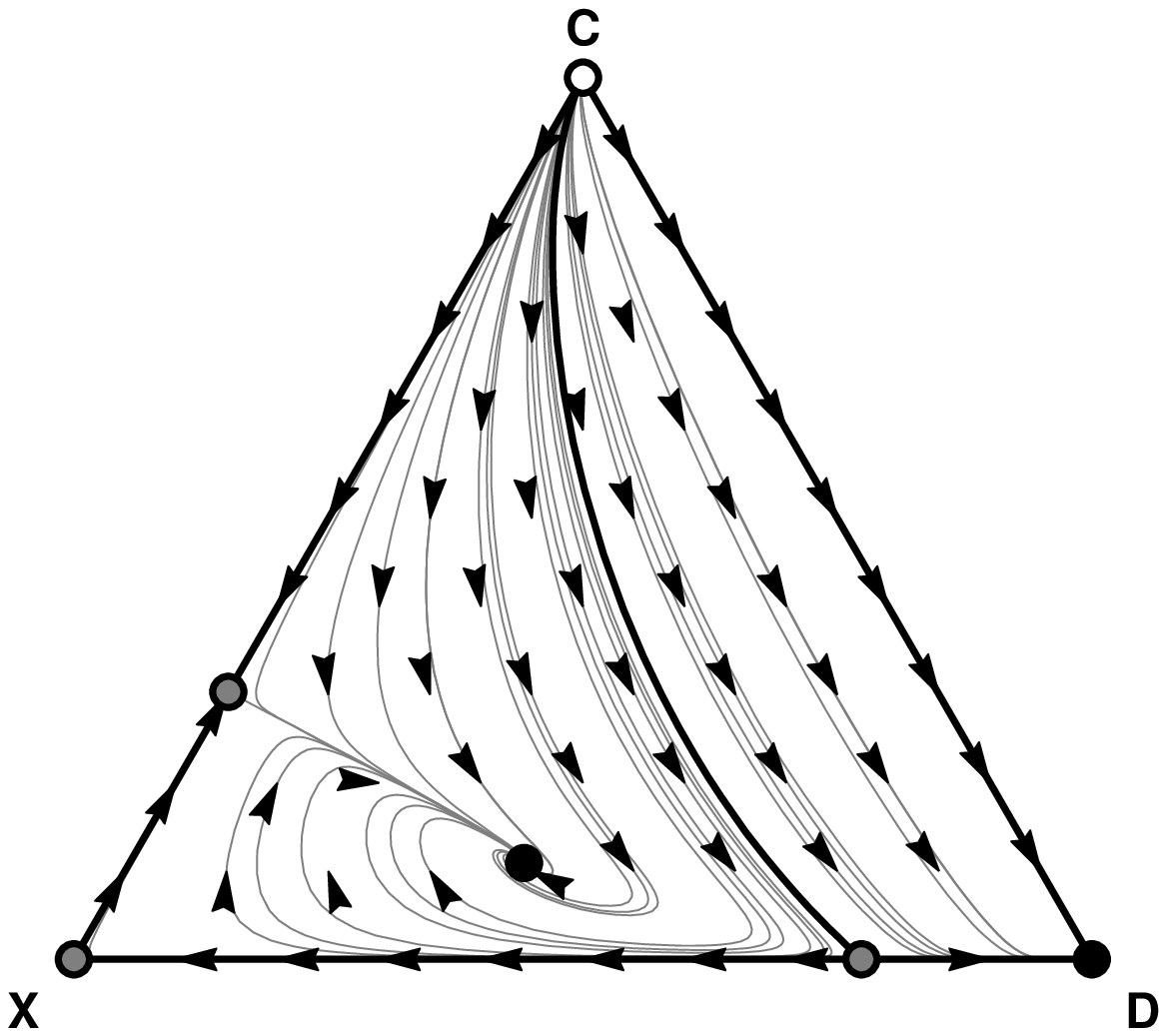}\hspace*{5mm}
\includegraphics[width=0.30\textwidth,clip]{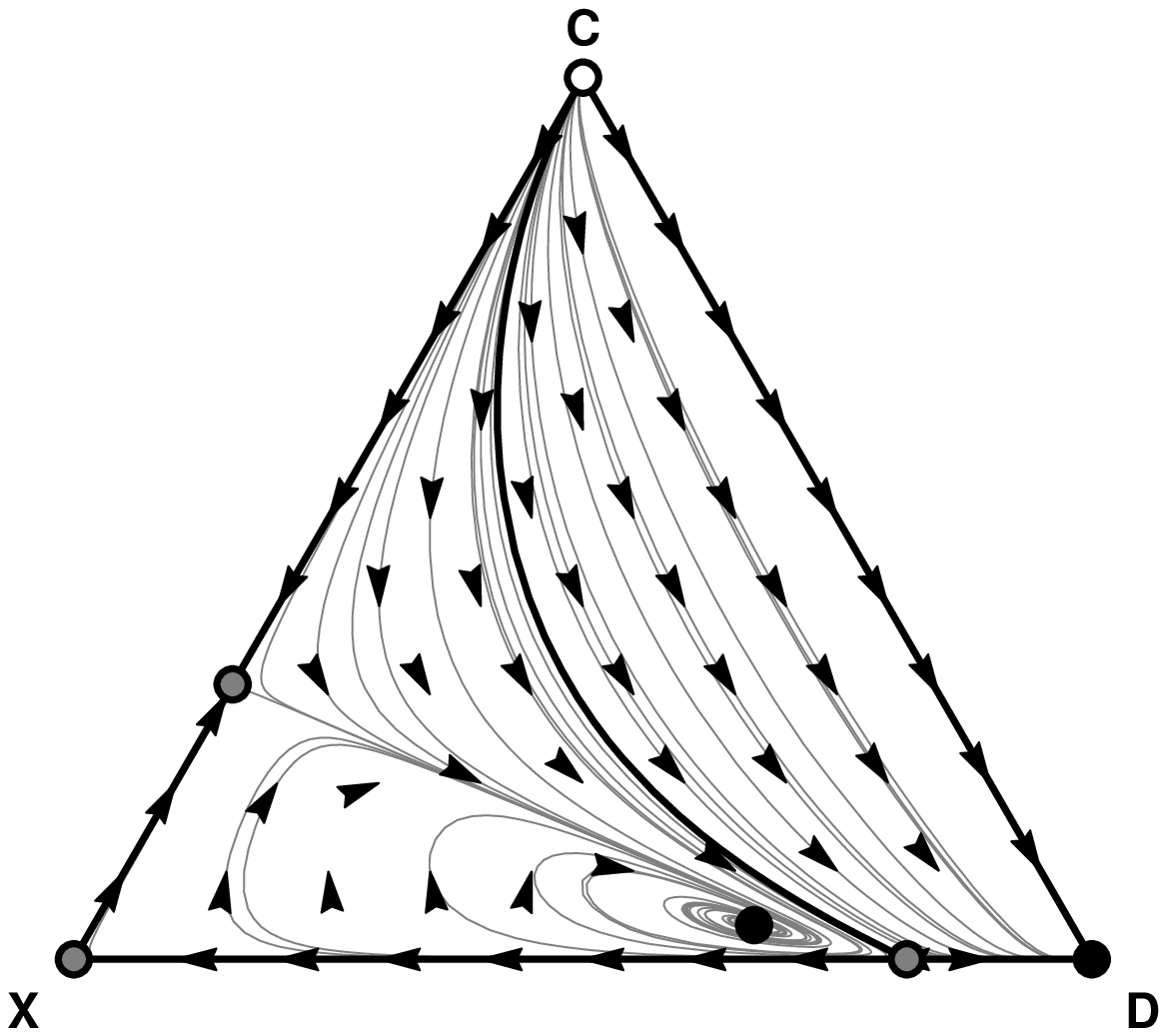}
\caption{Phase portraits of the replicator dynamics for 2-players IMPD games
with three strategies (C, D, and X) for different values
of $p$: $0.80$ (left), $0.83$ (middle), and $0.90$ (right). Other parameters:
$q=0.2$, $p_0=0.4$, $p_1=0.8$. Rest points marked in the plot 
can be repellors (white), saddle points (grey) or  attractors (black). In
all three cases the inner point as well as the $x_D=1$ point are the only
attractors of the system.}
\label{fig:2p-diff_p}
\end{figure*}
hand, there will be $(n+2)(n+1)/2$ such matrices displaying all possible
combinations of the three strategies (C, D, X). Obtaining the expressions
for them is of course straightforward, but doing it analytically for $n>2$
is out of question. Once the matrices are obtained computing the vector $\pi$
containing the $2^n$ stationary probabilities for each of the states simply
amounts again to solving the linear system $\pi=\pi M$, readily providing 
the payoffs for any strategy $i$ when confronted with any set $i_1,\dots,i_n$
of strategies of the $n-1$ opponents. The result can be cast in a tensor
$W=(W_{i,i_1,\dots,i_{n-1}})$. For a population composition $x$ the payoff
received by an individual of strategy $i$ will thus be
\begin{equation}
W_i(x)=\sum_{i_1,\dots,i_{n-1}=\text{C, D, X}}W_{i,i_1,\dots,i_{n-1}}x_{i_1}
\cdots x_{i_{n-1}},
\label{eq:payoffn}
\end{equation}
and the average payoff of the population will be
\begin{equation}
\overline{W}(x)=\sum_{i=\text{C, D, X}}x_iW_i(x).
\label{eq:averagepayoffn}
\end{equation}
Finally, the replicator dynamics is then given by
\begin{equation}
\dot x_i=x_i[W_i(x)-\overline{W}(x)].
\label{eq:replicatorn}
\end{equation}

Expression~\eqref{eq:payoffn} can be further simplified if we exploit
the symmetry implicit in public goods games, where the identity of the
players is not at all relevant, only the number of them using a
given strategy. This means that many payoffs are equal because
\begin{equation}
W_{i,i_1,\dots,i_{n-1}}=W_i(n_{\rm C},n_{\rm D},n_{\rm X}),
\label{eq:symmetricpayoffs}
\end{equation}
i.e., the payoff obtained by an $i$ strategist only depends on the
number $n_{\rm C}$ of cooperators, $n_{\rm D}$ of defectors, and
$n_{\rm X}$ of conditional cooperators ($n_{\rm C}+
n_{\rm D}+n_{\rm X}=n-1$) she is confronted to. Then
\begin{equation}
W_i(x)=\sum_{\substack{n_{\rm C}+n_{\rm D}+n_{\rm X}=n-1 \\
n_{\rm C},n_{\rm D},n_{\rm X}\ge 0}}\frac{(n-1)!}{n_{\rm C}!
n_{\rm D}!n_{\rm X}!}W_i(n_{\rm C},n_{\rm D},n_{\rm X})
x_{\rm C}^{n_{\rm C}}x_{\rm D}^{n_{\rm D}}x_{\rm X}^{n_{\rm X}}.
\label{eq:payoffnsimple}
\end{equation}

\begin{figure*}[t]
\centering
\includegraphics[width=0.30\textwidth,clip]{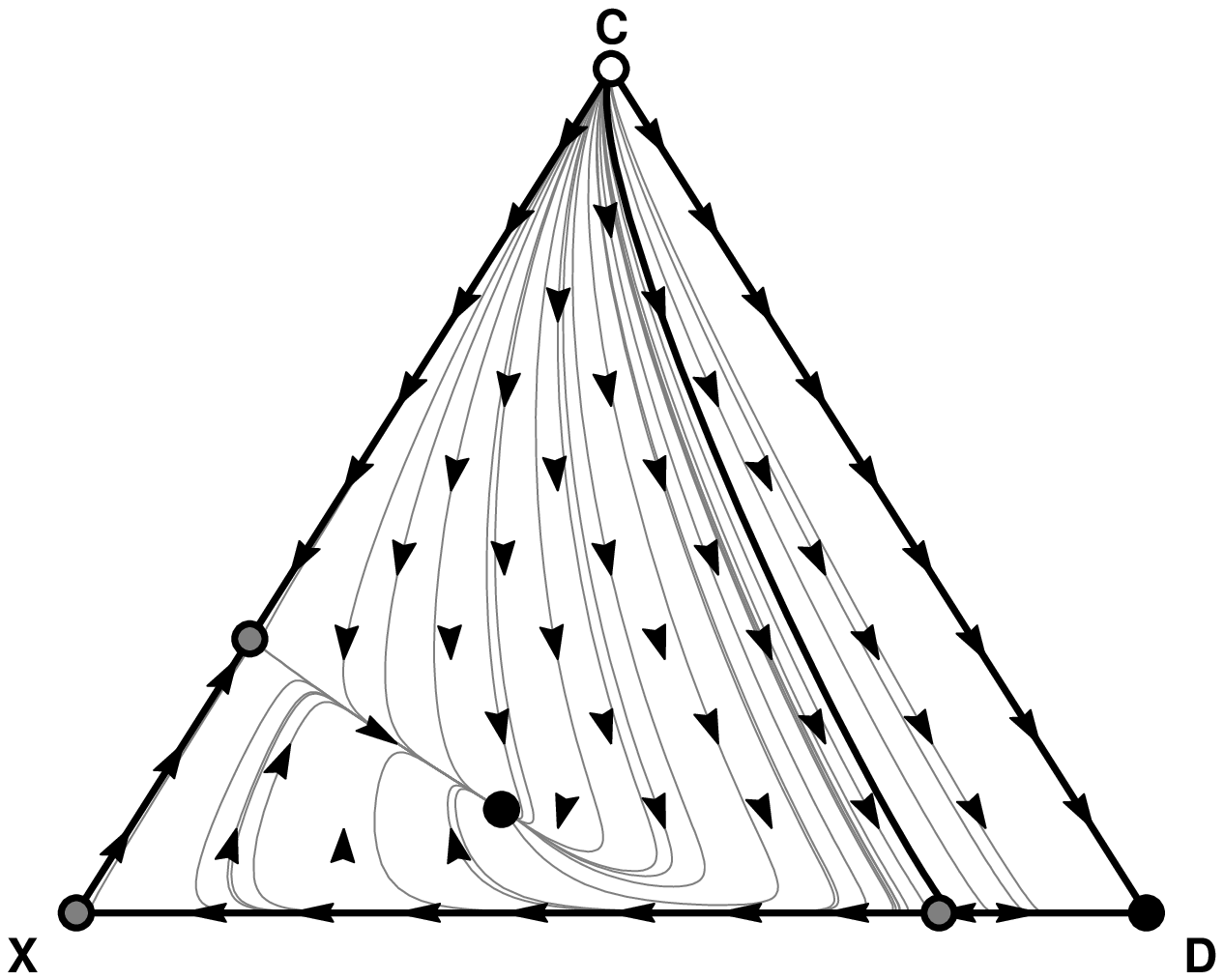}\hspace*{5mm}
\includegraphics[width=0.30\textwidth,clip]{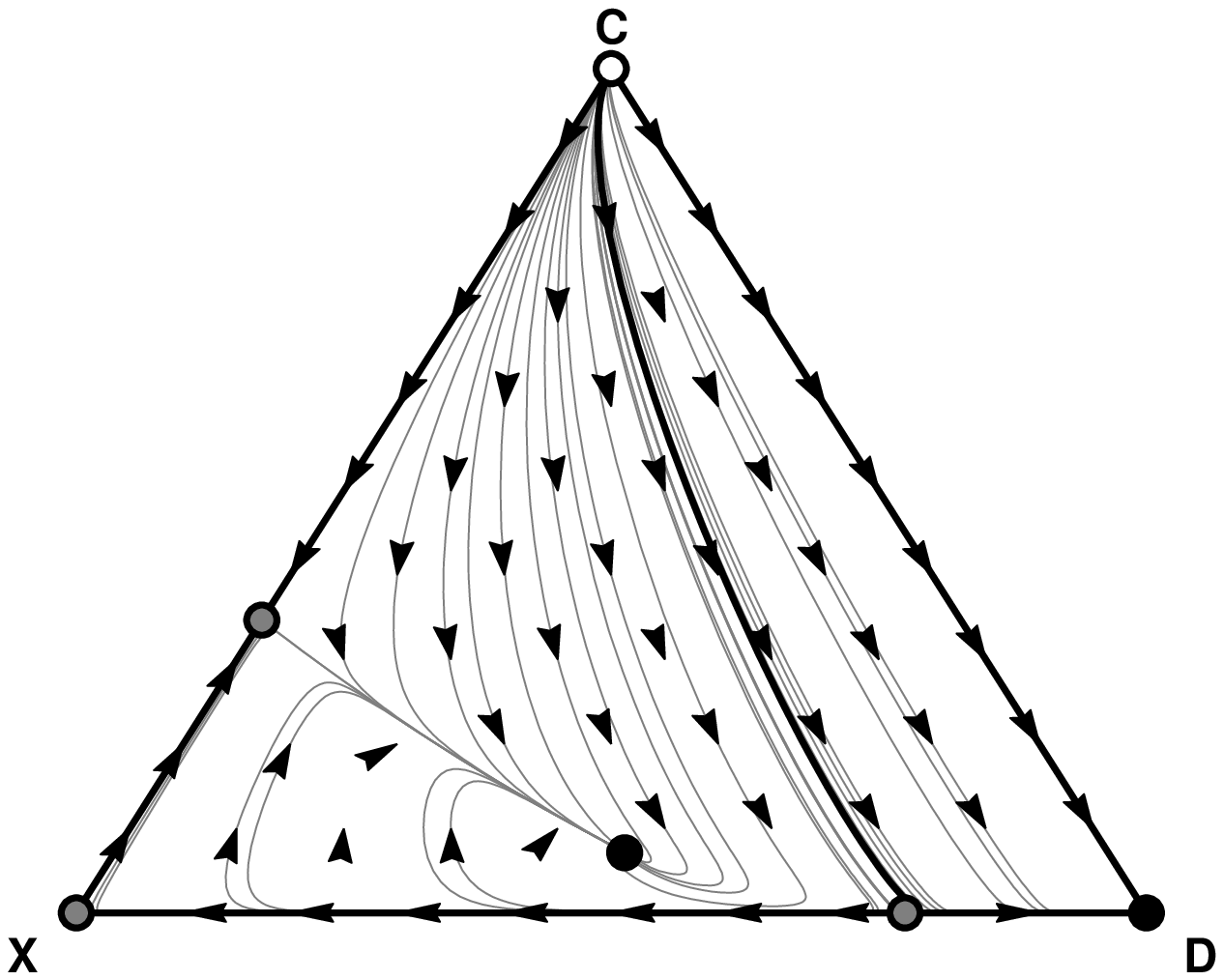}\hspace*{5mm}
\includegraphics[width=0.30\textwidth,clip]{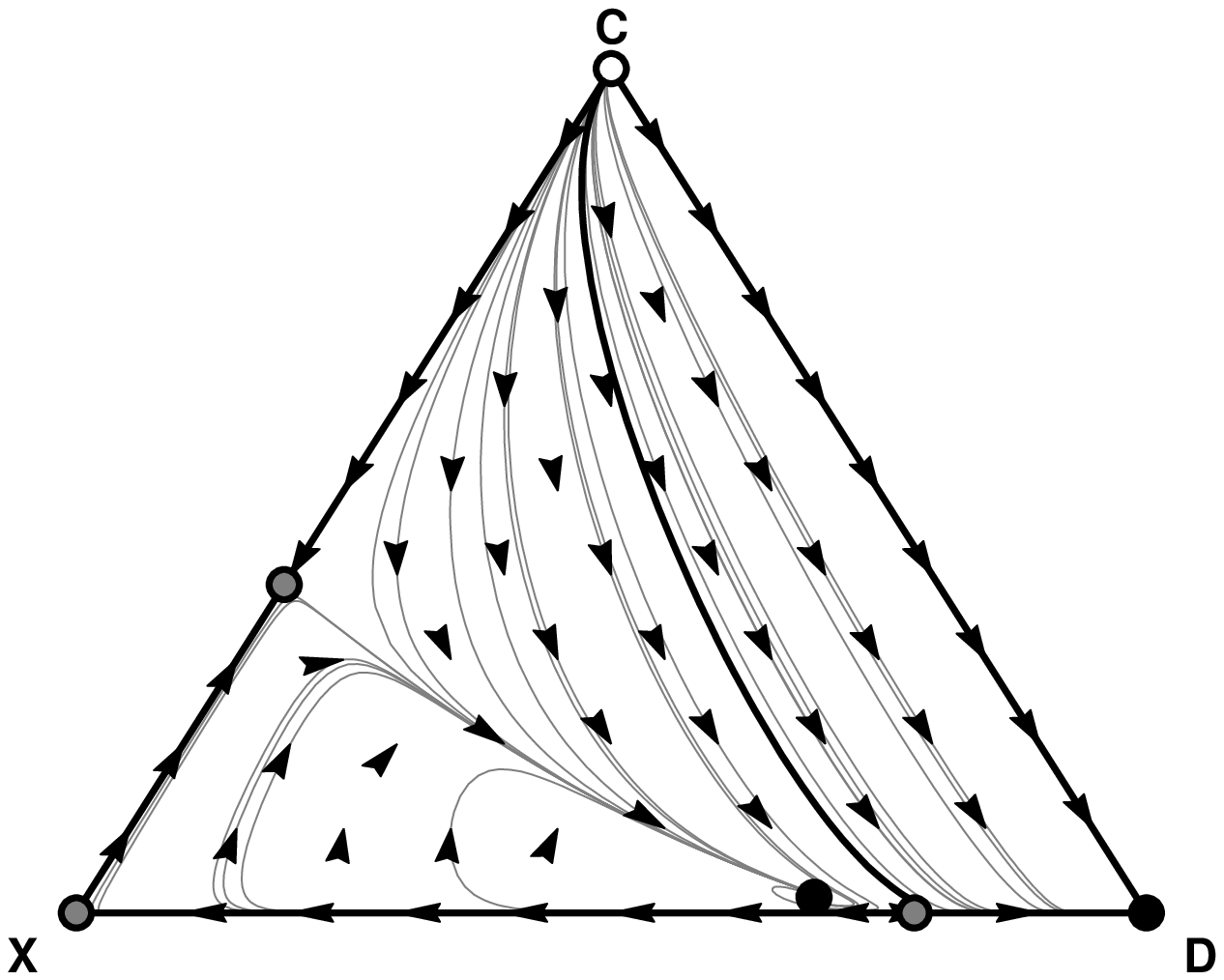}\hspace*{5mm}
\caption{Phase portraits of the replicator dynamics for 3-players IMPD games
with three strategies (C, D, and X) for different values
of $p$: $0.90$ (left), $0.92$ (middle), and $0.95$ (right). Other parameters:
$q=0.10$, $p_0=0.20$, $p_1=0.95$. Rest points marked in the plot 
can be repellors (white), saddle points (grey) or  attractors (black). In
all three cases the inner point as well as the $x_D=1$ point are the only
attractors of the system.}
\label{fig:3p-diff_p}
\end{figure*}

\begin{figure*}[t]
\centering
\includegraphics[width=0.30\textwidth,clip]{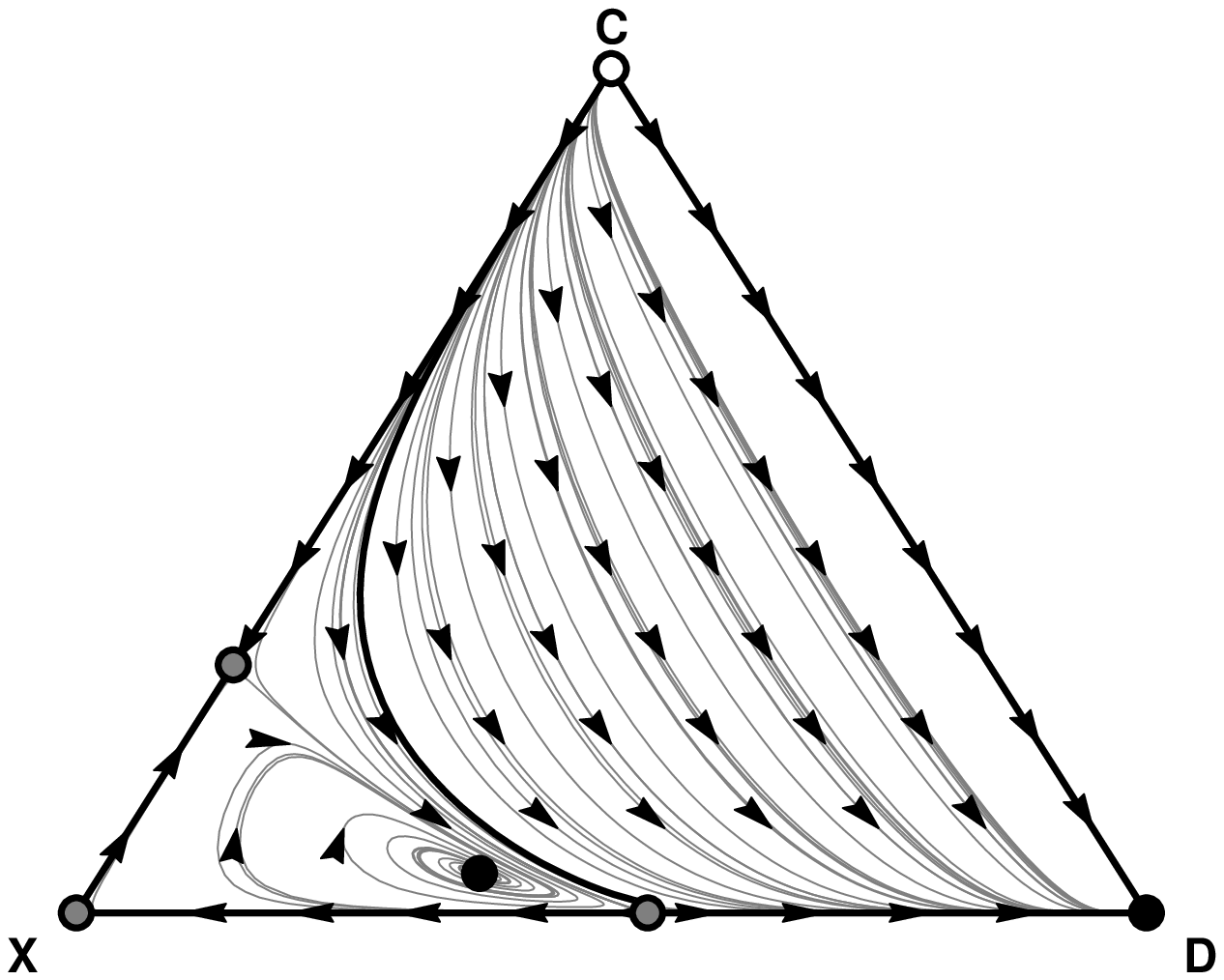}\hspace*{5mm}
\includegraphics[width=0.30\textwidth,clip]{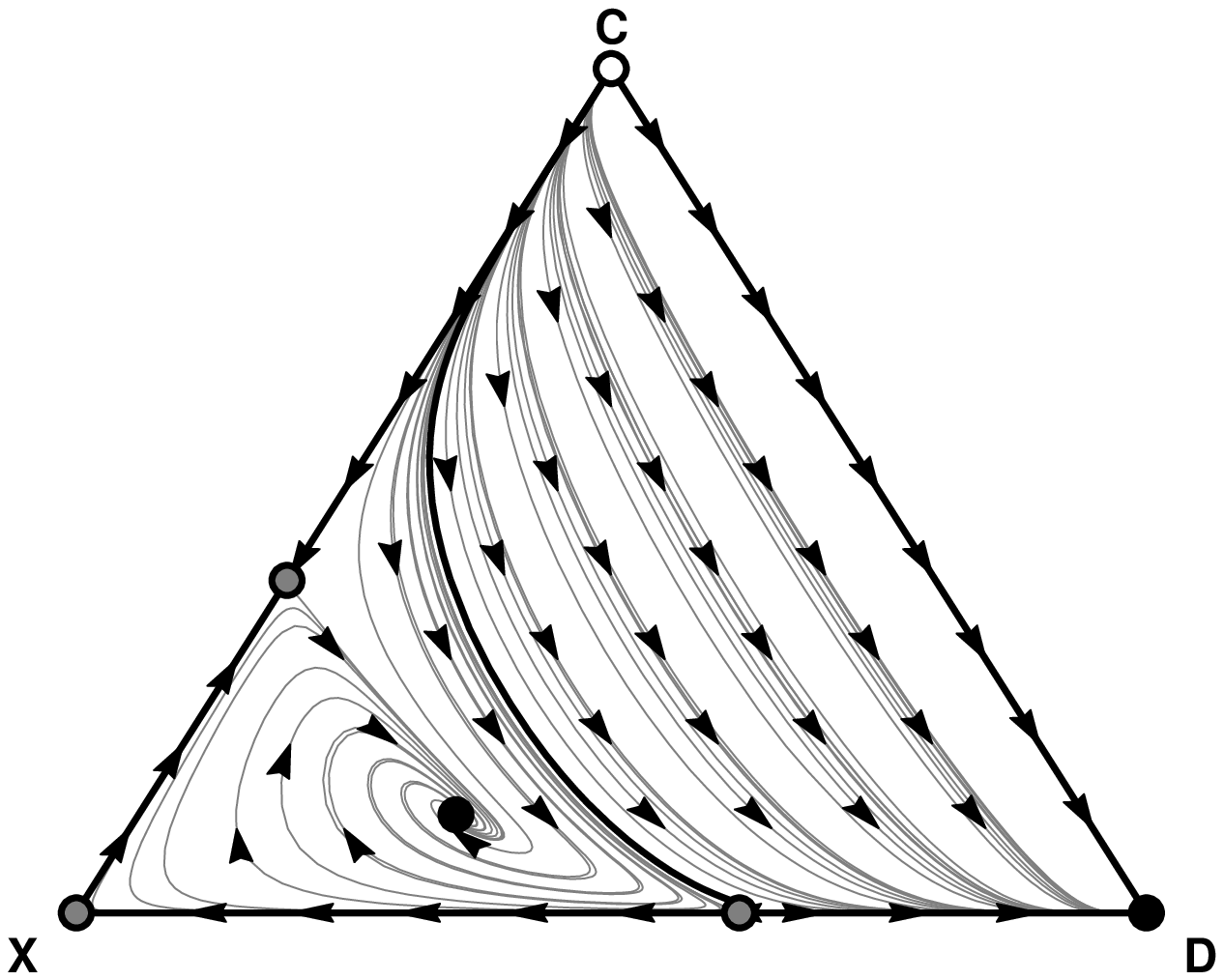}\hspace*{5mm}
\includegraphics[width=0.30\textwidth,clip]{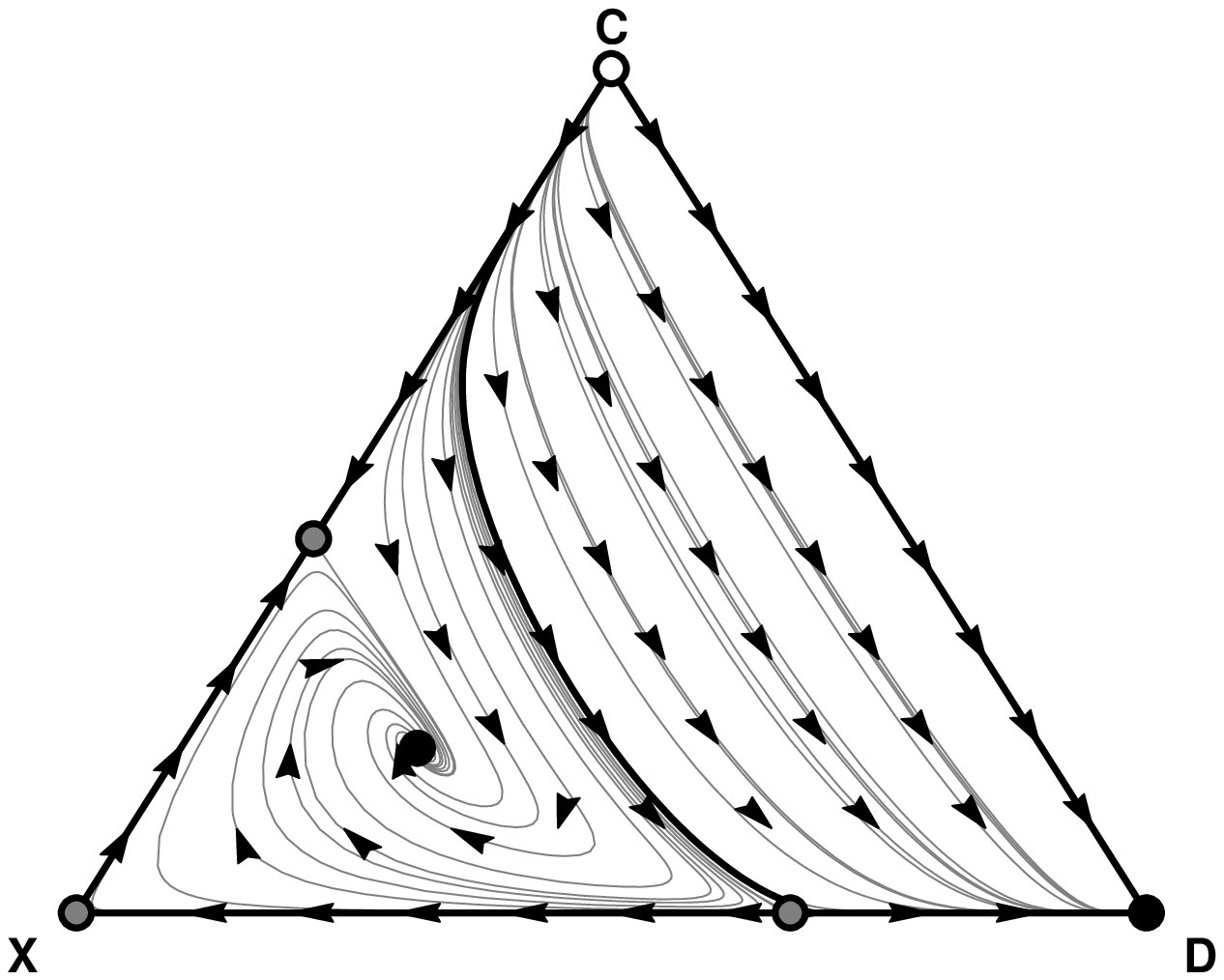}
\caption{Phase portraits of the replicator dynamics for 4-players IMPD games
with three strategies (C, D, and X) for different values
of $p_1$: $0.95$ (left), $0.97$ (middle), and $0.98$ (right). Other parameters:
$p=0.95$, $q=0.20$, $p_0=0.30$. Rest points marked in the plot 
can be repellors (white), saddle points (grey) or  attractors (black). In
all three cases the inner point as well as the $x_D=1$ point are the only
attractors of the system.}
\label{fig:4p-diff_p1}
\end{figure*}

As in Sec.\ 3, for the parameters obtained from the experiments there
is no interior point that describes the coexistence of the three
strategies. We subsequently proceeded as in the previous case and
tried to find ranges of parameters for which such an
interior point exists. It turns out that for 
groups of $n=3$ players sets of parameters can also be found where
the dynamics is similar to that for $n=2$ (see Figure~\ref{fig:3p-diff_p} for 
an example),
albeit the parameters for which this happens are a bit different ---but still
reasonably close to those of the experiments of \cite{grujic:2010}. As in
the two player case, the structure displayed in this figures turns out to
be extremely sensitive to variations in the parameters. Although we will not 
go into the details of those modifications here, we find it interesting to note that
Figure~\ref{fig:3p-diff_p} shows an evolution of the interior point with increasing
$p$ very similar to that for $n=2$ (cf.\ Figure~\ref{fig:2p-diff_p}), albeit with 
more drastic changes, indicating that the existence of an interior point
is less generic. For IMPDs with larger groups we find that, although
for groups of $n=4$ players it is still possible to find a Zeeman-like
phase map, one has to choose values for $p$ very close to one (meaning that
cooperators and defectors are nearly pure strategies) and on top of that
the region where this behavior can be obtained is extremely narrow. 
It can be clearly observed 
in Figure~\ref{fig:4p-diff_p1}, where several of these maps are shown for
different values of $p_1$, that variations of about $1\%$ noticeably displace
the location of the interior point. Importantly, it can be also observed from 
Figure~\ref{fig:3p-diff_p} and Figure~\ref{fig:4p-diff_p1} that
the basin of attraction of the interior point, when it exists, shrinks
upon increasing the number of players,
i.e., for $n=4$ the fraction of trajectories that end up in the 
D attractor is larger than those ending in the interior point. Finally, for the 
largest group size we could handle computationally, $n=5$ players, we have not
been able to find an interior point for any choice of parameters. It
turns out that the outcome of this game for $n\ge 5$ is
well represented by the large group limit $n\to\infty$, which unlike the
case of arbitrary but finite $n$, is amenable to analysis ---as we show in
the next section.

\section{Infinitely large groups}

Obtaining the payoffs \eqref{eq:symmetricpayoffs} amounts to finding
the stationary state of $(n+2)(n+1)/2$ Markov chains, each made of
$(n_{\text{C}}+1)\times (n_{\text{D}}+1)\times(n_{\text{X}}+1)$ states,
where $n_{\text{C}}+n_{\text{D}}+n_{\text{X}}=n$ defines the composition
of the $n$-player group. The size of the corresponding Markov matrices
grows as $n^3$, which makes it feasible to study groups even larger than
$n=5$ players. This will not be necessary though, because 
the resulting chain can be studied analytically in the limit $n\to\infty$,
which characterizes well the behavior of large groups.

To determine how a group with $n_{\text{C}}+n_{\text{D}}+n_{\text{X}}=n$
players of each type will respond in a given iteration of the prisoner's
dilemma we only need to record the vector $(k_{\text{C}},k_{\text{D}},
k_{\text{X}})$ whose components count how many players of each strategy
cooperate in a given round. Then the probability to observe the Markov
chain in a certain state given that in the previous round the state was
$(l_{\text{C}},l_{\text{D}},l_{\text{X}})$ is
\begin{equation}
\begin{split}
\Pr\big\{ &k_{\text{C}},k_{\text{D}},k_{\text{X}}|l_{\text{C}},l_{\text{D}},
l_{\text{X}}\big\}=\binom{n_{\text{C}}}{k_{\text{C}}}
\binom{n_{\text{D}}}{k_{\text{D}}}
p^{n_{\text{D}}-k_{\text{D}}+k_{\text{C}}}
(1-p)^{n_{\text{C}}-k_{\text{C}}+k_{\text{D}}} \\
&\times\sum_{j=0}^{k_{\text{X}}}\binom{l_{\text{X}}}{j}
\binom{n_{\text{X}}-l_{\text{X}}}{k_{\text{X}}-j}
p_C(x)^j[1-p_C(x)]^{l_{\text{X}}-j}\\
&\times
q^{k_{\text{X}}-j}(1-q)^{n_{\text{X}}-l_{\text{X}}-k_{\text{X}}+j},
\end{split}
\end{equation}
with the usual convention that $\binom{a}{b}=0$ for $b>a$ and where 
we have introduced the short-hand notation
\[
x\equiv\frac{l_{\text{C}}+l_{\text{D}}+l_{\text{X}}-1}{n-1}.
\]
Extracting
analytical information for finite $n$ from this matrix is not an easy task. However, 
let us focus on the limit $n\to\infty$. It is straightforward
to show that
\begin{equation}
\begin{split}
\E\left(\left.\frac{k_{\text{C}}}{n}\right|l_{\text{C}},l_{\text{D}},
l_{\text{X}}\right) &=p\frac{n_{\text{C}}}{n}, \\
\E\left(\left.\frac{k_{\text{D}}}{n}\right|l_{\text{C}},l_{\text{D}},
l_{\text{X}}\right) &=(1-p)\frac{n_{\text{D}}}{n}, \\
\E\left(\left.\frac{k_{\text{X}}}{n}\right|l_{\text{C}},l_{\text{D}},
l_{\text{X}}\right) &=p_C(x)\frac{l_{\text{X}}}{n}
+q\frac{n_{\text{X}}-l_{\text{X}}}{n},
\end{split}
\end{equation}
and
\begin{equation}
\begin{split}
\Var\left(\left.\frac{k_{\text{C}}}{n}\right|l_{\text{C}},l_{\text{D}},
l_{\text{X}}\right) =&p(1-p)\frac{n_{\text{C}}}{n^2}, \\
\Var\left(\left.\frac{k_{\text{D}}}{n}\right|l_{\text{C}},l_{\text{D}},
l_{\text{X}}\right) =&p(1-p)\frac{n_{\text{D}}}{n^2}, \\
\Var\left(\left.\frac{k_{\text{X}}}{n}\right|l_{\text{C}},l_{\text{D}},
l_{\text{X}}\right) =&p_C(x)[1-p_C(x)]\frac{l_{\text{X}}}{n^2} \\
&+q(1-q)\frac{n_{\text{X}}-l_{\text{X}}}{n^2}.
\end{split}
\end{equation}
Hence, introducing the random variable $r_i\equiv k_i/n_i$  and denoting
$x_i\equiv n_i/n$, in the limit
$n\to\infty$ the probability density of $r_i$ becomes a delta function
around $r_{\text{C}}=p$, $r_{\text{D}}=1-p$ and $r_{\text{X}}$,
this last quantity arising from the solution to the equation
\begin{equation}
r_{\text{X}}=\big\{p_0+(p_1-p_0)[px_{\text{C}}+(1-p)x_{\text{D}}
+r_{\text{X}}]\big\}r_{\text{X}}+ q(1-r_{\text{X}}).
\end{equation}
If $p_0=p_1$ this is a linear equation with solution $r_{\text{X}}=
q/(1-p_0+q)$. If $p_0\ne p_1$ it is a quadratic equation with two
solutions. The one that reduces to the solution found for $p_0=p_1$
is
\begin{align}
r_{\text{X}} &= \frac{2q}{\Delta+\sqrt{\Delta^2-4q(p_1-p_0)x_{\text{X}}}}, \\
\Delta &\equiv 1-p_0+q-(p_1-p_0)[px_{\text{C}}+(1-p)x_{\text{D}}].
\end{align}
Notice that $\Delta>0$ as long as $p_1>p_0>q$, as required.

Factors $r_i$ yield the asymptotic, stationary fraction
of cooperative actions among players of type $i$ in the group. Hence
the stationary level of cooperation is given by
\begin{equation}
\kappa\equiv px_{\text{C}}+(1-p)x_{\text{D}}+r_{\text{X}}x_{\text{X}},
\end{equation}
and the corresponding payoffs of the three type of players are
\begin{align}
W_{\text{C}}(x) =&p\kappa R+p(1-\kappa)S+(1-p)\kappa T+(1-p)(1-\kappa)P, \\
W_{\text{D}}(x) =&(1-p)\kappa R+(1-p)(1-\kappa)S+p\kappa T+p(1-\kappa)P, \\
W_{\text{X}}(x) =&r_{\text{X}}\kappa R+r_{\text{X}}(1-\kappa)S+
(1-r_{\text{X}})\kappa T +(1-r_{\text{X}})(1-\kappa)P. 
\end{align}

Notice that
\begin{equation}
W_{\text{D}}(x)-W_{\text{C}}(x)=(2p-1)[\kappa(T-R)+(1-\kappa)(P-S)],
\end{equation}
so as long as $p>1/2$ we have $W_{\text{D}}(x)>W_{\text{C}}(x)$ i.e.,
cooperators are always dominated by defector irrespective of the
composition of the population (provided $x_{\text{D}}>0$). This implies
that no interior point exists in the limit
$n\to\infty$, a property that suggests that the fact that 
we have not been able to locate
an interior point for $n=5$ is generic for larger values of $n$.

On the other hand,
\begin{align}
W_{\text{C}}(x)-W_{\text{X}}(x) &=(r_{\text{X}}-p)
[\kappa(T-R)+(1-\kappa)(P-S)],\\
W_{\text{D}}(x)-W_{\text{X}}(x) &=(r_{\text{X}}+p-1)
[\kappa(T-R)+(1-\kappa)(P-S)],
\end{align}
so any solution to $r_{\text{X}}=p$ ($r_{\text{X}}=1-p$) 
determines a rest point on the
$x_{\text{D}}=0$ ($x_{\text{C}}=0$) edge of the simplex. 
Taking the first equation and assuming
$x_{\text{D}}=0$ we obtain
\[
\left(\frac{2q}{p}-\Delta\right)^2=\Delta^2-4q(p_1-p_0)x_{\text{X}}.
\]
Upon simplification this
equation becomes
\[
q+p^2(x_{\text{C}}+x_{\text{X}})=p(1-p_0+q).
\]
Given that $x_{\text{C}}+x_{\text{X}}=1$ on the $x_{\text{D}}=0$ edge
of the simplex, it turns out that $r_{\text{X}}=p$ does not hold
for any point of this edge. A similar argument yields the same
result for $r_{\text{X}}=1-p$ on the $x_{\text{C}}=0$ edge of the
simplex (the equations are the same just replacing $p$ by $1-p$ and
$x_{\text{C}}$ by $x_{\text{D}}$).

We have thus established that, depending on the parameters $p_1>p_0>q$
and $p>1/2$, on the $x_{\text{D}}=0$ edge of the simplex either
$W_{\text{C}}(x)>W_{\text{X}}(x)$ or $W_{\text{C}}(x)<W_{\text{X}}(x)$
irrespective of the composition, and on the $x_{\text{C}}=0$ edge
of the simplex either $W_{\text{D}}(x)>W_{\text{X}}(x)$ or
$W_{\text{D}}(x)<W_{\text{X}}(x)$ irrespective of the composition.
In order to decide which one of the inequalities holds on each
edge we can set an arbitrary composition, namely $x_{\text{X}}=1$.
At this corner of the simplex
\begin{equation}
\begin{split}
r_{\text{X}} &=\frac{2q}{1-p_0+q+\sqrt{(1-p_0+q)^2-4q(p_1-p_0)}} \\[3mm]
&=\frac{1-p_0+q-\sqrt{(1-p_0-q)^2+4q(1-p_1)}}{2(p_1-p_0)}.
\end{split}
\end{equation}
Then $W_{\text{C}}(x)>W_{\text{X}}(x)$ on $x_{\text{D}}=0$ if, and only if,
\begin{equation}
\frac{1-p_0+q-\sqrt{(1-p_0-q)^2+4q(1-p_1)}}{2(p_1-p_0)}>p>\frac{1}{2},
\label{eq:WC>WX}
\end{equation}
and $W_{\text{D}}(x)>W_{\text{X}}(x)$ on $x_{\text{C}}=0$ if, and only if,
\begin{equation}
\frac{1-p_0+q-\sqrt{(1-p_0-q)^2+4q(1-p_1)}}{2(p_1-p_0)}>1-p.
\label{eq:WD>WX}
\end{equation}
Notice that if \eqref{eq:WC>WX} is true so is \eqref{eq:WD>WX}
(but the converse does not hold).

For \eqref{eq:WC>WX} to hold a necessary condition is that the
left-hand side is larger than $1/2$, a condition that boils down to
\[
\begin{split}
1-p_1+q &>\sqrt{(1-p_0-q)^2+4q(1-p_1)} \\
&=\sqrt{(1-p_1+q)^2+2(p_1-p_0)\left(1-q-\frac{p_1+p_0}{2}\right)}.
\end{split}
\]
As $p_1>p_0$, the only way that this can hold is if $q+(p_1+p_0)/2>1$.
When inequality~\eqref{eq:WC>WX} is satisfied,
D is an attractor, X is a repellor, and C a saddle point. Otherwise
C is a repellor (obviously, a sufficient condition for this to happen
is $q+(p_1+p_0)/2<1$). In this case D is an attractor and X a saddle point if
\eqref{eq:WD>WX} holds and viceversa if it does not. 
\begin{figure}[t]
\centering
\includegraphics[width=0.45\textwidth,clip]{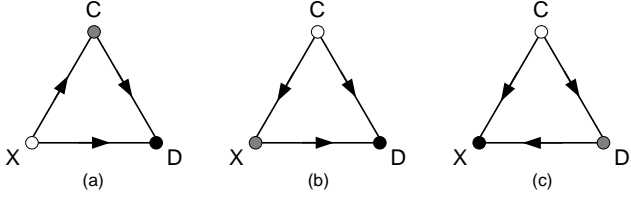}
\caption{The only three phase portraits of the replicator dynamics for 
IMPD games with three strategies (C, D, and X) played in infinitely
large groups. Rest points marked in the plot 
can be repellors (white), saddle points (grey) or  attractors (black). Map (a) appears if inequality~\eqref{eq:WC>WX} holds
(a necessary condition for this is $q+(p_1+p_0)/2>1$); map (b) appears if
inequality~\eqref{eq:WC>WX} does not hold but inequality~\eqref{eq:WD>WX}
does; map (c) appears if neither \eqref{eq:WC>WX} nor \eqref{eq:WD>WX}
hold (a sufficient condition for maps (b) and (c) to appear is
$q+(p_1+p_0)/2<1$).}
\label{fig:infinitegroups}
\end{figure}

A summary of our results for $n\to\infty$ is shown in the sketch 
of Figure~\ref{fig:infinitegroups}. As we can see from the plot, 
the main results are that there never exists an interior point,  that
homogeneous C populations are not stable, and that in two out
of three cases the final result of the dynamics is a homogeneous
$D$ population. Therefore, although there is a region of parameters
in which a homogeneous population of moody conditional cooperators 
is actually stable, we never
observe coexistence even of pairs of strategies.

\section{Discussion}

Motivated by the recent experimental work by  \cite{grujic:2010},
where conditional cooperation depending on the player's previous 
action was observed in a spatial prisoner's dilemma coexisting with
cooperation and defection, we have studied the replicator dynamics 
of the IMPD with these three strategies. The fact that the experimental
results indicated that all three strategies were getting on average 
the same payoff suggested that they were in equilibrium; on the other
hand, as the presence of a lattice had no significant consequences on
the level of cooperation, it seemed likely that the spatial game could 
be understood in terms of separate multiplayer games. 

Assuming a stylized version of the behaviors mentioned above, we have focused
on the problem of their coexistence in well-mixed populations, when they
interact in groups of $n\ge 2$ players through an IMPD. For $n=2$, 
in a region of parameters compatible
with those of the experiment we do find a mixed equilibrium in which
all three types of players coexist, and they do it in a proportion similar
to that found in the experiments. The phase portrait of the replicator dynamics
reproduces that of a three-strategies game introduced by \cite{zeeman:1980}.
However, upon increasing $n$, the region of parameters of this Zeeman-like
dynamics shrinks, and for $n=5$, the maximum size we could analyze with our
analytical approach, we could not find a mixed equilibrium
anymore.

Given that our Markov chain technique becomes computationally untraceable
for larger sizes, we have carried out a rigorous analysis of the replicator dynamics for this
game in the limit $n\to\infty$. The analysis reveals that in this limit, all
rest points other than the three corners of the simplex ---that can be found
for small $n$--- disappear. The dynamics in this limit is determined by
who beats who, depending on the parameters. Cooperators are always defeated
by defectors, but depending on the parameters, conditional cooperators are
displaced by any other strategy, or only by defectors, or they can 
displace the other two strategies.

Putting together our numerical results for small $n$ and our analytical
calculations for large $n$, we can conclude that
an imitative evolution like the one represented by replicator dynamics
cannot account for the coexistence of strategies observed in the experiments,
at least in groups as large as $n=9$ (the case of the experiment). The
reasons for this can be many. The most obvious one is that replicator
dynamics might not be what describes the evolution of strategies in human
subjects. In this regard, we have to make it clear that we are not studying 
the evolution of the players during the experiment, as it was shown by 
\cite{grujic:2010} that there is no learning. Our evolutionary approach would
apply to much longer time scales, i.e., these strategies would have arisen 
from interactions of human groups through history. It may then well be the 
case that this slower evolution of human behavior requires another approach
to its dynamics. By the same token, it might also occur that the typical 
number of iterations of the game is not very large, so the stationary 
probability density obtained from the Markov chains is not a good 
approximation to the observed behavior. All in all, it is clear that our 
analytical model might not be the most appropriate one to describe human
behavior on IMPDs. 

Nevertheless, another possible explanation for the discrepancy between 
our predictions and the coexistence of moody conditional cooperators with
the cooperator and defector strategists might come from bounded
rationality considerations. Thus, people may behave in a IMPD as 
though they were playing a (two-person) IPD with some kind of an 
``average'' opponent, something that can be reinforced by the computer
interface of the experiment that isolates the subjects from the other ones
with whom they interact. Such a heuristic decision making process might
be the result of cognitive biases or limitations, among which the inability
to deal with large numbers may be of relevance here \citep{kahneman:1982},
or else it could arise as an adaptation itself \citep{gigerenzer:2001}.
Whatever the underlying reason, the fact that for $n=2$ and $n=3$ players
we can easily find wide ranges of parameters for which the three 
strategies coexist and, furthermore, this coexistence 
have a large basin of attraction, suggests that the idea that people may 
be extrapolating their behavior to larger groups should at least be 
considered, and tested by suitably designed experiments. 

On the 
other hand, it should be borne in mind that the strategies reported by 
\cite{grujic:2010} are aggregate behaviors, as they attempted to classify
the actions of the player in a few archetypal types. Therefore, there may
actually be very many different moody conditional cooperators, defined
by different $p_0$, $p_1$ and $q$ parameters and
different propensities to cooperate (parameter $p$) among cooperators
and defectors. Alternatively players who were classified as conditional
cooperators might be using a totally different strategy, different for
every player, which aggregated would look like the conditional
cooperation detected in the experiment.
This is not included
anywhere in our replicator dynamics. It is certainly possible that considering
several different subclasses of the strategy X in the replicator dynamics 
might actually provide an explanation for coexistence in larger groups. 
However, the corresponding calculations become very much involved, 
and whether this variability can sustain mixed equilibria
is an interesting question that remains out of the scope of this work.

As a final remark, we would like to stress that, notwithstanding the 
issue that the agreement between our results and the experiments 
is problematic, this study proves that, under replicator dynamics, 
even for $n\to\infty$
our work predicts the dominance of moody conditional cooperators for
certain regions of parameters. It is important to realize that this type of
strategy had not been considered prior to the experimental observation, 
and as we now see it can successfully take over the entire population 
even from defection when playing an IMPD. This suggests that this 
or similar strategies may actually be more widespread than this simple
case as they might also be the best ones in related games, such as 
the public goods game. It would be worth widening the scope of this
work by analyzing the possible appearance of this conditional cooperators
who are influenced by their own mood in other contexts, both theoretically 
and experimentally. In this regard, an explanation of the evolutionary
origin of moody conditional cooperators would be a particularly important,
albeit rather difficult goal.

\section*{Acknowledgments}

This work was supported in part by MICINN (Spain) through grant MOSAICO,
by ERA-NET Complexity-Net RESINEE, and by Comunidad de Madrid (Spain)
through grant MODELICO-CM.

\appendix

\section{Zeeman's game}
\label{sec:appendix}

\begin{figure*}
\centering
\includegraphics[width=0.80\textwidth,clip]{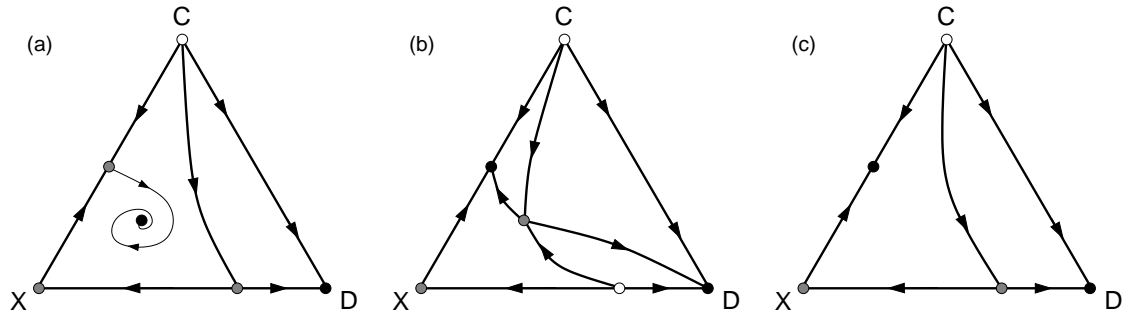}
\caption{Three different phase maps that can emerge from the replicator
dynamics for a Zeeman game (c.f.~Eq.~\eqref{eq:Zeemanpayoff}). There
may (pannels (a) and (b)) or may not be (pannel (c)) an interior point,
and it may be an attractor (pannel (a)) or a saddle point (pannel (b)).
Circles mark the rest points and arrows indicate the direction of the
flux. White circle denote unstable rest points, grey circles denote
saddle points, and black circles denote stable points.}
\label{fig:zeeman}
\end{figure*}

\cite{zeeman:1980} analyzed the evolutionary dynamics of three strategies games.
Appart from the well known rock-paper-scissors \citep{hofbauer:1998} he identified
a game with the canonical payoff matrix for the strategies C, D and X, given by
\begin{equation}
\begin{pmatrix}
0 & -a_2 & b_1 \\
b_2 & 0 & -a_3 \\
a_1 & -b_3 & 0
\end{pmatrix},
\label{eq:Zeemanpayoff}
\end{equation}
where all coefficients are positive. Any $3\times 3$ payoff matrix can be
transformed into a zero diagonal one because the replicator equation remains
invariant if the same constant is substracted from every element of one of
its columns \citep{hofbauer:1998}.
The coefficients of the payoff matrix \eqref{eq:Zeemanpayoff} represent
the payoff an invader gets when it invades a homogeneous population. Thus
a D or X individual invading a homogeneous C population will get $b_2$ or
$a_1$, respectively. As both are positive a homogeneous C population
is unstable. Similarly a C or X individual invading a homogeneous D
population will get $-a_2$ or $-b_3$, respectively. Therefore a homogeneous
D population is uninvadable (hence stable). As for a C or a D individual
invading a homogeneous X population, it will obtain $b_1$ or $-a_3$,
respectively. It is therefore a saddle point because it cannot be invaded
by $D$ individuals but it can by $C$ individuals.

This simple analysis fixes the flux of the dynamics at the boundary of the
simplex (Figure~\ref{fig:zeeman}). It also implies the existence of two rest
points on the boundary of the simplex: one on the D--X edge and another one
on the C--X edge. These points are given by
\begin{equation}
\left(0,\frac{a_3}{a_3+b_3},\frac{b_3}{a_3+b_3}\right), \qquad
\left(\frac{b_1}{a_1+b_1},0,\frac{a_1}{a_1+b_1}\right).
\label{eq:middlepoints}
\end{equation}
Besides, an interior rest point $(y_{\text{C}},y_{\text{D}},y_{\text{X}})
/(y_{\text{C}}+y_{\text{D}}+y_{\text{X}})$, with coordinates
\begin{equation}
\begin{split}
y_{\text{C}} &=b_3(a_3+b_1)-a_2a_3, \\
y_{\text{D}} &=b_1b_2-a_1(b_1+a_3), \\
y_{\text{X}} &=a_1a_2+b_2b_3-a_2b_2,
\end{split}
\label{eq:interior}
\end{equation}
appears provided all three components have the same sign (Figure~\ref{fig:zeeman}(a)).
Component $y_D$ is proportional to the difference between the payoff of the
population at the C--X mixed equilibrium and the payoff of a D invader. When
it is negative the C--X rest point becomes a saddle and the interior point
is an attractor (the situation depicted in Figure~\ref{fig:zeeman}(a)). When it
is positive a D individual cannot invade the C--X equilibrium, which then
becomes an attractor and the interior point becomes a repellor (this is
illustrated in Figure~\ref{fig:zeeman}(b)). If no interior point exists the
behavior will be as plotted in Figure~\ref{fig:zeeman}(c) \citep{zeeman:1980}. 


\end{document}